\documentclass[12pt]{article}

\usepackage{amsmath,amssymb,amsfonts,amsthm}
\usepackage{graphicx}
\usepackage{cite}
\usepackage[all]{xy}
\usepackage[toc,page]{appendix}
\usepackage{hyperref}
\hypersetup{colorlinks=true,  citecolor=red, linkcolor=blue}

\allowdisplaybreaks[3]

\newcounter{propositiona}
\newcommand{\propositiona}[1]{\refstepcounter{propositiona}
\noindent
\textbf{Proposition \thepropositiona.}\, {\it #1}}
\newcounter{definitiona}
\newcommand{\definitiona}[1]{\refstepcounter{definitiona}
\noindent
\textbf{Definition \thedefinitiona.}\, #1}
\newcounter{remarka}
\newcommand{\remarka}[1]{\refstepcounter{remarka}
\noindent
\textbf{Remark \theremarka.}\, #1}
\newcounter{examplea}
\newcommand{\examplea}[1]{\refstepcounter{examplea}
\noindent
\textbf{Example \theexamplea.}\, #1}
\newcounter{lemmaa}

\newcounter{theorema}
\newcommand{\theorema}[1]{\refstepcounter{theorema}
\noindent
\textbf{Theorem\, \thetheorema.}\, {\it #1}}
\newcounter{corollarya}

\renewcommand{\thefootnote}{\alph{footnote}}

\title{Invariant Reduction for Partial Differential Equations. II: The General Framework}

\author{ \renewcommand{\thefootnote}{\alph{footnote}}
Kostya Druzhkov\footnotemark[1],~~Alexei Cheviakov\footnotemark[2]\vspace{0.5cm}\\
\small $^{\rm a,b}$\emph{Department of Mathematics and Statistics, University of Saskatchewan, Saskatoon, Canada}\vspace{0.2cm}\\
}

\begin{document}

\footnotetext[1]{Corresponding author. Electronic mail: konstantin.druzhkov@gmail.com}
\footnotetext[2]{Electronic mail: shevyakov@math.usask.ca}

\maketitle \numberwithin{equation}{section}
\renewcommand{\thefootnote}{\arabic{footnote}}

\begin{abstract}
For a system of partial differential equations (PDEs) $F = 0$ admitting a local (point, contact, or higher) symmetry $X$ with the characteristic $\varphi$, invariant solutions satisfy the reduced system $F = \varphi = 0$. We propose a framework that allows, for every $X$-invariant conservation law, presymplectic structure, variational principle, or another geometric structure of the given PDE system $F = 0$, to systematically calculate its corresponding reduced form that describes the corresponding structure for the reduced system $F = \varphi = 0$. In particular, we show in what way Noether's theorem holding for the given PDE system is inherited by the reduced PDE system. We consider several detailed examples, including cases of point and higher symmetry invariance. 
The proposed framework is directly applicable to a wide range of PDE models, including complex PDE systems of contemporary interest arising across disciplines, where symmetry reduction is essential for analysis and simulation, as well as to integrable, Lagrangian, and gauge systems.

\end{abstract}

{\bf Keywords:} Invariant solutions, Lie symmetries, higher symmetries, conservation laws, presymplectic structures, Liouville integrability

\section{Introduction}

Symmetry reductions of partial differential equations (PDEs) are a fundamental tool in nonlinear science, allowing the transformation of complex multidimensional problems into simpler, lower-dimensional forms. For instance, if a system exhibits invariance under geometric symmetries such as translations, rotations, or scalings -- or more general Lie point symmetries -- solutions that remain unchanged under these symmetry actions can be systematically sought \cite{ovsiannikov1982group, bibintro30, Olver, bluman2010applications}. 
Symmetry reductions have been widely applied to construct exact solutions of nonlinear PDEs that are both mathematically significant and physically relevant (see, e.g., Refs.~\cite{cantwell2002introduction, bluman2010applications, bogoyavlenskij2000mhd, cheviakov2015fully, dierkes2020new}). Furthermore, reduced PDE systems often exhibit enhanced analytical structures, such as additional symmetries, conservation laws, or even integrability properties, which are absent in the original model (see, e.g., Refs.~\cite{Olver, cheviakov2024analytical, kelbin2013new} and references therein). 

One of the significant advantages of symmetry-based methods is their algorithmic nature. Lie point symmetries, along with their generalizations, including contact, higher\footnote{Also commonly referred to as \emph{higher-order} or \emph{generalized} symmetries.}, and nonlocal symmetries, can be systematically identified using symbolic computation tools. This makes symmetry methods applicable to a wide range of PDE models in diverse fields of science and engineering (see, e.g., Refs.~\cite{Olver, bluman2010applications} and references therein).

The vast majority of results in the literature dealing with symmetry reductions and their applications employs point symmetries, because invariance under point symmetries often yields explicit DE systems with fewer independent variables. In the current series of papers, we extend this approach to work seamlessly for point, contact, and higher symmetries. 

In the first part of this study \cite{InvRedI}, a reduction procedure was described that utilized a local symmetry and a symmetry-invariant conservation law for PDE systems with two independent variables. This method enabled the algorithmic computation of constants of motion for symmetry-invariant solutions. However, from a dynamical and structural point of view, reductions of conservation laws alone are often not sufficient for studying exact solutions or integrability properties of the systems describing invariant motions. The current work extends the reduction procedure of Ref.~\cite{InvRedI} to other structures by revealing its homological underpinnings. Specifically, for systems of PDEs with symmetries, we analyze symmetry-invariant geometric structures and propose the concept of their reduction as elements of cohomology groups of cochain complexes. This approach generalizes methods for symmetry-invariant conservation law reduction developed in Refs.~\cite{sjoberg2007double, sjoberg2009double, bokhari2010generalization} and partially overlaps with the results of Ref.~\cite{anco2020symmetry} in the case of reductions of conservation laws based on a single point symmetry.
The reduction framework proposed here is broadly applicable to any PDE system with symmetries\footnote{We also assume the presence of suitable invariant structures and the compatibility of the DE system describing invariant solutions.}. It is based on the following simple observation.

\medskip\noindent\textbf{The main idea.} For a system of differential equations $F = 0$ and its evolutionary symmetry $X$ with a characteristic $\varphi$, invariant solutions satisfy the system
\begin{align}
F = 0\,,\qquad \varphi = 0\,.
\label{introsys}
\end{align}
The symmetry $X$ vanishes on this system. Suppose $\omega$ is an equivalence class of differential forms that represents an $X$-invariant element of a cohomology group of some cochain complex with a differential~$\partial$. Then for some class $\vartheta$ and the Lie derivative $\mathcal{L}_X \omega$, we have
\begin{align*}
\mathcal{L}_X \omega = \partial \vartheta.
\end{align*}
Thus, the restriction of $\partial \vartheta$ to the system for $X$-invariant solutions is zero, and the restriction of $\vartheta$ is a cocycle of the corresponding complex for~\eqref{introsys}, provided that such a complex exists. The cohomology class of this cocycle is the desired reduction of the cohomology class of $\omega$. In particular, $\omega$ can represent a conservation law, a presymplectic structure, or an internal Lagrangian~\cite{DRUZHKOV2023104848} of $F = 0$.

\medskip

A related reduction approach was proposed in Ref.~\cite{anderson1997symmetry}, where symmetry reductions are formulated in terms of flows of vector fields on jet manifolds and the symmetry-invariant part of the variational bicomplex~\cite{tsujishita1982variation}. In its original formulation, however, this mechanism is not directly applicable to higher symmetries. Nonetheless, it allows one to treat suitable groups of point transformations in the multidimensional case without requiring their solvability, whereas the reduction mechanism developed here admits multi-reduction by non-commutative symmetry algebras only under certain conditions.

The paper is organized as follows. In Section~\ref{Basnot}, we introduce notation and recall some concepts from the geometry of differential equations. Section~\ref{InvRedmain} introduces the reduction mechanism for the first page of the Vinogradov $\mathcal{C}$-spectral sequence and for the stationary action principle. It establishes some relationships between $\mathcal{C}$-spectral sequences of a system of PDEs and systems that describe its symmetry-invariant solutions. In addition, it clarifies some challenges related to multi-reduction, and proposes a version of Noether's theorem adapted to the PDE system satisfied by invariant solutions.
Section~\ref{Computal} contains computational algorithms to calculate reductions of conservation laws and presymplectic structures of systems of evolution equations. In Section~\ref{Examplessec}, we provide examples of reduction of conservation laws, presymplectic structures, and the stationary action principle. In Section~\ref{sec:red:cl}, we calculate a reduction of a conservation law of a (1+2)-dimensional nonlinear diffusion equation with respect to a scaling symmetry; the reduced conservation law has a form of a vanishing total curl. Further, for the Calogero–Bogoyavlenskii–Schiff breaking soliton equation, we examine reduction of two conservation laws under a higher symmetry. In this example, the reduced conservation laws give rise to a single constant of invariant motion. In Section~\ref{sec:red:pre}, we present three examples of reduction of presymplectic structures: for a nonlinear breaking wave equation $u_{tt} = (1 + u_x^2)u_{xx}$ \cite{cheviakov2016one, cheviakov2020invariant, mcadam2025nonlinear}, the potential Kaup–Boussinesq system (e.g., Ref.~\cite{cheviakov2024analytical}), and the cotangent covering of the $r^{\rm th}$ dispersionless Dym equation (see, e.g., Refs.~\cite{blaszak2002classical, pavlov2003integrable, morozov2009contact, ovsienko2010bi}). In Section~\ref{sec:red:var}, we illustrate the framework of reduction of variational principles using the examples of the nonlinear breaking wave equation and the nonlinear Schr\"odinger equation with respect to a point and a higher symmetry, respectively. We discuss the inheritance of Liouville integrability via invariant reduction of the nonlinear Schr\"odinger equation. The straightforward computational nature of the examples is aimed at offsetting the abstractness of our approach.

We use the Einstein summation notation throughout this paper and consider only smooth functions of the class $C^{\infty}$.

\section{Basic notation and definitions \label{Basnot}}

Let us introduce notation and briefly recall basic facts from the geometry of differential equations.

\subsection{Jets}

We now briefly review the notion of jet bundles and related structures. For details, see, e.g.,~\cite{VinKr}.

\vspace{1ex}

Let $\pi\colon E^{n+m}\to M^n$ be a locally trivial smooth vector bundle over a smooth manifold $M^n$. Suppose $U\subset M$ is a coordinate neighborhood such that the bundle $\pi$
becomes trivial over $U$. Choose local coordinates $x^1$, \ldots, $x^n$ in $U$ and $u^1$, \ldots, $u^m$
along the fibers of $\pi$ over $U$. In these coordinates, a section $\sigma\in \Gamma(\pi)$ has the form of a smooth vector function
\begin{align*}
\sigma\colon\qquad u^1 = \sigma^1(x^1, \ldots, x^n)\,,\qquad \ldots\,, \qquad u^m = \sigma^m(x^1, \ldots, x^n)\,.
\end{align*}
Two sections $\sigma_1, \sigma_2 \in \Gamma(\pi)$ define the same $k$-jet ($k = 0, 1, \ldots, \infty$) at a point $x_0\in U$, if for $i = 1, \ldots, m$, the functions $\sigma^i_1$ and $\sigma^i_2$ have the same $k$-degree Taylor polynomials at $x_0$. We denote by $[\sigma]^k_{x_0}$ the $k$-jet of $\sigma\in \Gamma(\pi)$ at $x_0\in M$. The set $J^{k}(\pi)$ of all $k$-jets of sections of $\pi$ is naturally endowed with a smooth manifold structure.
On $J^{k}(\pi)$, one can introduce adapted local coordinates, given by
\begin{align}
x^i([\sigma]^{k}_{x_0}) = x^i_0\,,\qquad u^i_{\alpha}([\sigma]^{k}_{x_0}) = \dfrac{\partial^{|\alpha|} \sigma^i}{(\partial x^1)^{\alpha_1}\ldots (\partial x^n)^{\alpha_n}}(x_0)\,,\qquad |\alpha| \leqslant k\,.
\label{adaptedcoordinates}
\end{align}
Here $\alpha$ is a multi-index, $|\alpha| = \alpha_1 + \ldots + \alpha_n$.
It is convenient to treat $\alpha$ as a formal sum of the form $\alpha = \alpha_1 x^1 + \ldots + \alpha_n x^n = \alpha_i x^i$, where all $\alpha_i$ are non-negative integers. In what follows, we consider only adapted local coordinates on $J^{k}(\pi)$.

Note that $u^i_{\alpha}$ are functions on some open subset of $J^{k}(\pi)$. We shall call them \emph{derivatives} due to~\eqref{adaptedcoordinates}, but we do not imply that they are functions of independent variables; indeed, the interpretation of local coordinates on $J^{k}(\pi)$ as functions of $x^1$, $\ldots$, $x^n$ would lead to a confusion.

\vspace{0.5ex}
\noindent
\textbf{Functions.}
The projections $\pi_{\infty, \,k}\colon J^{\infty}(\pi)\to J^k(\pi)$, $[\sigma]^{\infty}_{x_0}\mapsto [\sigma]^{k}_{x_0}$ allow one to define the algebra (over $\mathbb{R}$) of smooth functions on the infinite jet space $J^{\infty}(\pi)$
$$
\mathcal{F}(\pi) = \bigcup_{k\geqslant 0} \pi_{\infty,\hspace{0.2ex} k}^{\hspace{0.1ex} *}\, C^{\infty}(J^k(\pi))\,.
$$

\noindent
\textbf{Cartan distribution.} The main structure on jet manifolds is the Cartan distribution.
Using adapted local coordinates on $J^{\infty}(\pi)$, one can introduce the total derivatives
$$
D_{x^i} = \partial_{x^i} + u^k_{\alpha + x^i}\partial_{u^k_{\alpha}}\qquad\quad i = 1, \ldots, n.
$$
The planes of the Cartan distribution $\mathcal{C}$ on $J^{\infty}(\pi)$ are spanned by the total derivatives. It is convenient to interpret tangent vectors/vector fields on $J^{\infty}(\pi)$ in terms of derivations of $\mathcal{F}(\pi)$.

\vspace{0.5ex}
\noindent
\textbf{Cartan forms.} The Cartan distribution $\mathcal{C}$ determines the ideal $\mathcal{C}\Lambda^*(\pi)$
of the algebra
$$
\Lambda^*(\pi) = \bigcup_{k\geqslant 0} \pi_{\infty,\hspace{0.2ex} k}^{\hspace{0.1ex} *}\, \Lambda^*(J^k(\pi))
$$
of differential forms on $J^{\infty}(\pi)$.
The ideal $\mathcal{C}\Lambda^*(\pi)$ is generated by Cartan (or contact) forms, i.e., differential forms that vanish on all planes of the Cartan distribution $\mathcal{C}$.
A Cartan $1$-form $\omega\in\mathcal{C}\Lambda^1(\pi)$ can be written as a finite sum
$$
\omega = \omega_i^{\alpha}\theta^i_{\alpha}\,,\qquad\ \theta^i_{\alpha} = du^i_{\alpha} - u^i_{\alpha + x^k}dx^k
$$
in adapted local coordinates. The coefficients $\omega_i^{\alpha}$ are smooth functions of adapted coordinates.

\vspace{0.5ex}
\noindent
\textbf{Infinitesimal symmetries.} Denote by $\pi_k$ the projection $\pi_k\colon J^k(\pi)\to M$, $\pi_k\colon [\sigma]^k_{x_0}\mapsto x_0$.
Smooth sections of the pullback bundles $\pi^*_{k}(\pi)\colon \pi^*_{k}(E)\to J^k(\pi)$ naturally determine sections of the pullback $\pi_{\infty}^*(\pi)\colon \pi^*_{\infty}(E)\to J^{\infty}(\pi)$ by means of the projections $\pi_{\infty,\hspace{0.2ex} k}$. We denote by $\varkappa(\pi)$ the $\mathcal{F}(\pi)$-module of such sections of $\pi_{\infty}^*(\pi)$. Each section $\varphi\in \varkappa(\pi)$ gives rise to a corresponding evolutionary vector field $E_{\varphi}$ on $J^{\infty}(\pi)$. In adapted coordinates,
$$
E_{\varphi} = D_{\alpha}(\varphi^i)\partial_{u^i_{\alpha}}\,,
$$
where $\varphi^1$, \ldots, $\varphi^m$ are components of $\varphi$, $D_{\alpha}$ denotes the composition $D_{x^1}^{\ \alpha_1}\circ\ldots\circ D_{x^n}^{\ \alpha_n}$.
Evolutionary vector fields are evolutionary symmetries of $J^{\infty}(\pi)$. In particular, $\mathcal{L}_{E_{\varphi}}\,\mathcal{C}\Lambda^*(\pi)\subset \mathcal{C}\Lambda^*(\pi)$. Here $\mathcal{L}_{E_{\varphi}}$ is the corresponding Lie derivative. Elements of $\varkappa(\pi)$ are characteristics of symmetries of $J^{\infty}(\pi)$. An infinitesimal symmetry of $J^{\infty}(\pi)$ is a sum of an evolutionary vector field and a trivial symmetry, i.e., a combination of the total derivatives $D_{x^1}$, $\ldots$, $D_{x^n}$.

\vspace{0.5ex}
\noindent
\textbf{Horizontal forms.}
Cartan forms allow one to consider the modules of horizontal $k$-forms
$$
\Lambda^k_h(\pi) = \Lambda^k(\pi)/\mathcal{C}\Lambda^k(\pi)\,.
$$
The de Rham differential $d$ induces the horizontal differential $d_h\colon \Lambda^k_h(\pi)\to \Lambda^{k+1}_h(\pi)$.
The infinite jet bundle $\pi_{\infty}\colon J^{\infty}(\pi) \to M$ admits the decomposition
$$
\Lambda^1(\pi) = \mathcal{C}\Lambda^1(\pi) \oplus \mathcal{F}(\pi)\!\cdot\!\pi^*_{\infty}(\Lambda^1(M))\,.
$$
We identify the module of horizontal $k$-forms $\Lambda^k_h(\pi)$ with $\mathcal{F}(\pi)\cdot \pi^*_{\infty}(\Lambda^k(M))$.

In adapted local coordinates, elements of $\mathcal{F}(\pi)\cdot \pi^*_{\infty}(\Lambda^k(M))$ are generated by the differentials $dx^1, \ldots, dx^n$, while $d_h = dx^i\wedge D_{x^i}$. For example,
$$
d_h (\xi_j dx^j) = dx^i\wedge D_{x^i}(\xi_j) dx^j = D_{x^i}(\xi_j)dx^i\wedge dx^j\,.
$$

\noindent
\textbf{Euler operator.} Let $\widehat{\varkappa}(\pi)$ be the adjoint module
\begin{align*}
\widehat{\varkappa}(\pi) = \mathrm{Hom}_{\mathcal{F}(\pi)}(\varkappa(\pi), \Lambda^n_h(\pi))\,.
\end{align*}
Denote by $\mathrm{E}$ the Euler operator (variational derivative), $\mathrm{E}\colon \Lambda^n_h(\pi)\to \widehat{\varkappa}(\pi)$.
In adapted local coordinates, for $L = \lambda \, dx^1\wedge\ldots\wedge dx^n$ and $\varphi\in\varkappa(\pi)$, we have
\begin{align*}
&\mathrm{E}(L)\colon\varphi \mapsto \langle \mathrm{E}(L), \varphi\rangle = \dfrac{\delta \lambda}{\delta u^i}\,\varphi^i dx^1\wedge\ldots\wedge dx^n\,, \qquad \dfrac{\delta \lambda}{\delta u^i} = \sum_{\alpha} (-1)^{|\alpha|}D_{\alpha}\Big(\dfrac{\partial \lambda}{\partial u^i_{\alpha}}\Big)\,.
\end{align*}
Here $\langle \cdot, \cdot \rangle$ denotes the natural pairing between a module and its adjoint.

\subsection{Differential equations \label{SectionDiffEq}}

Let $\zeta \colon E_1\to M$ be a locally trivial smooth vector bundle over the same base as $\pi$. Smooth sections of the pullbacks $\pi^*_{r}(\zeta)$ determine a module of sections of the pullback $\pi^*_{\infty}(\zeta)\colon \pi^*_{\infty}(E_1)\to J^{\infty}(\pi)$. We denote it by $P(\pi)$. Any $F\in P(\pi)$ can be considered a (generally, nonlinear) differential operator $\Gamma(\pi)\to \Gamma(\zeta)$. Then $F = 0$ is a differential equation.
By its \emph{infinite prolongation} we mean the set of formal solutions $\mathcal{E}\subset J^{\infty}(\pi)$ defined by the infinite system of equations
\begin{align*}
\mathcal{E}\colon\qquad D_{\alpha}(F^i) = 0\,,\qquad |\alpha| \geqslant 0\,.
\end{align*}
Here $F^i$ are components of $F$ in adapted coordinates. We denote $\pi_{\mathcal{E}} = \pi_{\infty}|_{\mathcal{E}}$ and assume that $\pi_{\mathcal{E}}\colon \mathcal{E}\to M$ is surjective.

\vspace{1ex}
\remarka{We do not require that the number of equations of the form $F^i = 0$ coincides with the number of dependent variables $m$.}

\vspace{1ex}
\noindent
\textbf{Regularity conditions.} We say that $\mathcal{E}\subset J^{\infty}(\pi)$ is \emph{regular} if the following condition holds. A function $f\in \mathcal{F}(\pi)$ vanishes on $\mathcal{E}$ if and only if there is a differential operator $\Delta\colon P(\pi)\to \mathcal{F}(\pi)$ of the form $\Delta_i^{\alpha} D_{\alpha}$ (total differential operator) such that $f = \Delta(F)$. Here, for some integer $k$, the components $\Delta_i^{\alpha}$ with $|\alpha|\leqslant k$ may depend on the independent variables $x^i$, the dependent variables $u^i$, and derivatives up to order $k$, while $\Delta_i^{\alpha} = 0$ for $|\alpha| > k$.

Unless otherwise stated, we assume that all systems considered in Section~\ref{InvRedmain} are regular. We also assume that the de Rham cohomology groups $H^i_{dR}$ of all considered systems are trivial for~$i > 0$.

We say that $\mathcal{E}$ is a \emph{manifold} if it is representable\footnote{More precisely, if the algebra $\Lambda^*(\mathcal{E})$, defined below, coincides with the corresponding direct limit. Most systems arising in mathematical physics are manifolds, including all systems in extended Kovalevskaya form and arbitrary formally integrable systems (see, e.g., Ref.~\cite{guneysu2017profinite}).} as the inverse limit of some tower 
\begin{align*}
\xymatrix{
\ldots \ar[r] & \mathcal{E}_{1} \ar[r] & \mathcal{E}_0 \ar[r] & M,
}
\end{align*}
where all $\mathcal{E}_k$ are finite-dimensional smooth manifolds, all maps are surjective submersions, and the composition $\mathcal{E}\to \mathcal{E}_0\to M$ is $\pi_{\mathcal{E}}$.

\vspace{0.5ex}
\noindent
\textbf{Functions.}
By $\mathcal{F}(\mathcal{E})$ we denote the algebra of smooth functions on $\mathcal{E}$,
$$
\mathcal{F}(\mathcal{E}) = \mathcal{F}(\pi)|_{\mathcal{E}} = \mathcal{F}(\pi)/I\,.
$$
Here $I$ denotes the ideal of the system $\mathcal{E}\subset J^{\infty}(\pi)$, $I = \{f\in \mathcal{F}(\pi)\, \colon\, f|_{\mathcal{E}} = 0\}$. Tangent vectors/vector fields on $\mathcal{E}$ are defined in terms of derivations of the algebra $\mathcal{F}(\mathcal{E})$.

\vspace{0.5ex}
\noindent
\textbf{Cartan forms.} The ideal $\mathcal{C}\Lambda^*(\pi)\subset \Lambda^*(\pi)$ gives rise to the corresponding ideal $\mathcal{C}\Lambda^*(\mathcal{E})$ of the algebra\footnote{Note that if $\mathcal{E}$ is a manifold, then every vector field can be contracted with any differential form on it.} $\Lambda^*(\mathcal{E}) = \Lambda^*(\pi)/(I\cdot \Lambda^*(\pi) + \Lambda^*(\pi)\wedge dI)$ of differential forms on $\mathcal{E}$. The Cartan plane at a point of $\mathcal{E}$ is spanned by tangent vectors that annihilate all Cartan $1$-forms.

\vspace{0.5ex}
\noindent
\textbf{Solutions.} A section $\sigma\colon M\to \mathcal{E}$ of $\pi_{\mathcal{E}}$ is a \emph{solution} of the equation $\pi_{\mathcal{E}}$ if $\sigma^*(\mathcal{C}\Lambda^1(\mathcal{E})) = 0$. 
We consider only smooth sections of $\pi_{\mathcal{E}}$, i.e., such that $\sigma^*(\mathcal{F}(\mathcal{E}))\subset C^{\infty}(M)$.

\vspace{0.5ex}
\noindent
\textbf{Infinitesimal symmetries.} A \emph{symmetry} (more precisely, an infinitesimal symmetry) of an infinitely prolonged system of equations $\mathcal{E}$ is a vector field $X$ on $\mathcal{E}$ (a derivation of $\mathcal{F}(\mathcal{E})$) that preserves the Cartan distribution: $[X, \mathcal{C}D(\mathcal{E})]\subset \mathcal{C}D(\mathcal{E})$, where $\mathcal{C}D(\mathcal{E})$ denotes the module of \emph{Cartan derivations}, i.e., vector fields on $\mathcal{E}$ whose vectors lie in the respective planes of the Cartan distribution. Two symmetries are equivalent if they differ by a Cartan derivation (a trivial symmetry). One can say that, locally, trivial symmetries of a regular system are combinations of the total derivatives $\,\overline{\!D}_{x^i} = D_{x^i}|_{\mathcal{E}}$, $i = 1, \ldots, n$.

If $\varphi\in \varkappa(\pi)$ is a characteristic such that $E_{\varphi}$ is tangent to $\mathcal{E}$ (i.e., $E_{\varphi}(I)\subset I$), then the restriction $E_{\varphi}|_{\mathcal{E}}\colon \mathcal{F}(\mathcal{E})\to \mathcal{F}(\mathcal{E})$ is a symmetry of $\mathcal{E}$ (less formally, $E_{\varphi}$ can also be called a symmetry of $\mathcal{E}$).
For any symmetry $X$ of a regular system $\mathcal{E}\subset J^{\infty}(\pi)$, there exists $\varphi \in \varkappa(\pi)$ such that $X$ is equivalent to the restriction $E_{\varphi}|_{\mathcal{E}}$. In the regular case, symmetries of $\mathcal{E}\subset J^{\infty}(\pi)$ are in one-to-one correspondence with elements of the kernel of the linearization operator $l_{\mathcal{E}} = l_F|_{\mathcal{E}}\colon \varkappa(\mathcal{E})\to P(\mathcal{E})$, where $l_F\colon \varkappa(\pi)\to P(\pi)$, $\varphi \mapsto E_{\varphi}(F)$, $l_F(\varphi)^i = E_{\varphi}(F^i)$, and
$$
\varkappa(\mathcal{E}) = \varkappa(\pi)/I\cdot \varkappa(\pi)\,,\qquad P(\mathcal{E}) = P(\pi)/I\cdot P(\pi)\,.
$$

\vspace{0.5ex}
\noindent
\textbf{$\mathcal{C}$-spectral sequence.} For $p \geqslant 1$, the ideals $\mathcal{C}^p\Lambda^*(\mathcal{E}) = \mathcal{C}^p\Lambda^*(\pi)|_{\mathcal{E}}$ are stable with respect to the de Rham differential, $d(\mathcal{C}^p\Lambda^*(\mathcal{E})) \subset \mathcal{C}^p\Lambda^*(\mathcal{E})$.
Then the de Rham complex $\Lambda^{\bullet}(\mathcal{E})$ admits the filtration
$$
\Lambda^{\bullet}(\mathcal{E})\supset \mathcal{C}\Lambda^{\bullet}(\mathcal{E})\supset \mathcal{C}^2\Lambda^{\bullet}(\mathcal{E})\supset \mathcal{C}^3\Lambda^{\bullet}(\mathcal{E})\supset \ldots
$$
The corresponding spectral sequence $(E^{\hspace{0.1ex} p,\hspace{0.2ex} q}_r(\mathcal{E}), d^{\hspace{0.1ex} p,\hspace{0.2ex} q}_r)$ is the Vinogradov $\mathcal{C}$-spectral sequence~\cite{VINOGRADOV198441, VinKr}.
Here $\mathcal{C}^{k+1}\Lambda^k(\mathcal{E}) = 0$, $E^{\hspace{0.1ex} p, \hspace{0.2ex} q}_0(\mathcal{E}) = \mathcal{C}^p\Lambda^{p+q}(\mathcal{E})/\mathcal{C}^{p+1}\Lambda^{p+q}(\mathcal{E})$. All differentials $d_r^{\hspace{0.1ex} p,\hspace{0.2ex} q}$ are induced by the de Rham differential $d$,
$$
d_r^{\hspace{0.1ex} p,\hspace{0.2ex} q}\colon E^{\hspace{0.1ex} p, \hspace{0.2ex} q}_r(\mathcal{E}) \to E^{\hspace{0.1ex} p+r, \hspace{0.2ex} q+1-r}_r(\mathcal{E})\,,\qquad E^{\hspace{0.1ex} p, \hspace{0.2ex} q}_{r+1}(\mathcal{E}) = \ker d_r^{\hspace{0.1ex} p,\hspace{0.2ex} q}/ \mathrm{im}\, d_r^{\hspace{0.1ex} p-r,\hspace{0.2ex} q+r-1}\qquad \text{for}\quad r\geqslant 0\,.
$$
We also use the notation $d_r$ where it does not lead to confusion.

In particular, $\mathcal{C}$-spectral sequence allows one to define variational $k$-forms, conservation laws, and presymplectic structures of differential equations. A \emph{variational $k$-form} of $\mathcal{E}$ is an element of the group $E^{\hspace{0.1ex} k,\hspace{0.2ex} n-1}_1(\mathcal{E})$. A \emph{conservation law} of $\mathcal{E}$ is a variational $0$-form, i.e., an element of the group $E^{\hspace{0.1ex} 0,\hspace{0.2ex} n-1}_1(\mathcal{E})$.
A \emph{presymplectic structure} of $\mathcal{E}$ is a $d_1$-closed variational $2$-form, i.e., an element of the kernel of the differential
$$
d_1^{\hspace{0.2ex} 2,\hspace{0.2ex} n-1}\colon E^{\,2,\,n-1}_1(\mathcal{E})\to E^{\,3,\,n-1}_1(\mathcal{E}).
$$

In the regular case, each element of $E^{\hspace{0.1ex} p, \hspace{0.2ex} q}_0(\mathcal{E})$ has a unique representative in the restriction of $\mathcal{C}^p\Lambda^p(\pi)\wedge \pi_{\infty}^*(\Lambda^q(M))$.
To make the description less abstract, we identify elements of $E^{\hspace{0.1ex} p, \hspace{0.2ex} q}_0(\mathcal{E})$ with their representatives of this form. This allows introducing the vertical (or Cartan) differential
$$
d_v\colon E^{\hspace{0.1ex} p, \hspace{0.2ex} q}_0(\mathcal{E})\to E^{\hspace{0.1ex} p+1, \hspace{0.2ex} q}_0(\mathcal{E})\,,\qquad d_v = d - d_0\,.
$$
Here $d_v\circ d_0 = - d_0 \circ d_v$, and $d_v\circ d_v = 0$. Locally, on a regular $\mathcal{E}$, the differential $d_0$ takes the form
$$
d_0 = dx^i\wedge \mathcal{L}_{\,\overline{\!D}_{x^i}}\,,\qquad \,\overline{\!D}_{x^i} = D_{x^i}|_{\mathcal{E}}\,.
$$
If $\omega\in E^{\hspace{0.1ex} p, \hspace{0.2ex} q}_0(\mathcal{E})$ represents an element $\Omega\in E^{\hspace{0.1ex} p, \hspace{0.2ex} q}_1(\mathcal{E})$, then $d_v\hspace{0.1ex} \omega = d\omega$ represents $d_1\Omega$.

\vspace{0.5ex}
\noindent
\textbf{Internal Lagrangian formalism.} The Lagrangian formalism  can be encoded in terms of the intrinsic geometry of PDEs via internal Lagrangians \cite{DRUZHKOV2023104848, DRUZHKOV2024105143}. The result is the Hamiltonian formalism without time~\cite{DRUZHKOV2024ATMP}.

The de Rham differential $d$ induces the differentials in
\begin{align*}
0\rightarrow \mathcal{F}(\mathcal{E})\to \Lambda^1(\mathcal{E})\to \dfrac{\Lambda^2(\mathcal{E})}{\mathcal{C}^2\Lambda^2(\mathcal{E})}\to \ldots\to \dfrac{\Lambda^{n-1}(\mathcal{E})}{\mathcal{C}^2\Lambda^{n-1}(\mathcal{E})}\to \dfrac{\Lambda^{n}(\mathcal{E})}{\mathcal{C}^2\Lambda^{n}(\mathcal{E})}\to \dfrac{\Lambda^{n+1}(\mathcal{E})}{\mathcal{C}^2\Lambda^{n+1}(\mathcal{E})}\to 0
\end{align*}
Let us introduce the following notation
$$
\widetilde{E}_0^{\hspace{0.1ex} 0,\hspace{0.2ex} k}(\mathcal{E}) = \Lambda^{k+1}(\mathcal{E})/\mathcal{C}^2\Lambda^{k+1}(\mathcal{E})\,,\quad
\tilde{d}^{\hspace{0.2ex} 0,\hspace{0.2ex} k}_0\colon \widetilde{E}_0^{\hspace{0.1ex} 0,\hspace{0.2ex} k}(\mathcal{E})\to \widetilde{E}_0^{\hspace{0.1ex} 0,\hspace{0.2ex} k+1}(\mathcal{E})\,,\quad \widetilde{E}_1^{\hspace{0.1ex} 0,\hspace{0.2ex} k}(\mathcal{E}) = \ker \tilde{d}^{\hspace{0.2ex} 0,\hspace{0.2ex} k}_0/\mathrm{im}\, \tilde{d}^{\hspace{0.2ex} 0,\hspace{0.2ex} k-1}_0\,.
$$
The de Rham differential $d$ induces the differentials
$$
\tilde{d}^{\hspace{0.3ex} 0,\hspace{0.2ex} k}_1\colon \widetilde{E}_1^{\hspace{0.1ex} 0,\hspace{0.2ex} k}(\mathcal{E})\to E^{\hspace{0.1ex}2,\hspace{0.2ex} k}_1(\mathcal{E})\,.
$$
\emph{We identify each group $\widetilde{E}_1^{\hspace{0.1ex} 0,\hspace{0.2ex} k}(\mathcal{E})$ with its canonically isomorphic group}
\begin{align}
\widetilde{E}_1^{\hspace{0.1ex} 0,\hspace{0.2ex} k}(\mathcal{E}) = \dfrac{\{l\in \Lambda^{k+1}(\mathcal{E})\hspace{0.1ex} \colon \ dl\in \mathcal{C}^2\Lambda^{k+2}(\mathcal{E})\}}{\mathcal{C}^2\Lambda^{k+1}(\mathcal{E}) + d(\Lambda^{k}(\mathcal{E}))}\,.
\label{canonisom}
\end{align}
Elements of the group $\widetilde{E}_1^{\hspace{0.1ex} 0,\hspace{0.2ex} n-1}(\mathcal{E})$ are \emph{internal Lagrangians} of $\mathcal{E}$. The differential $\tilde{d}^{\hspace{0.3ex} 0,\hspace{0.2ex} n-1}_1$ maps them to presymplectic structures of $\mathcal{E}$.

If the variational derivative $\mathrm{E}(L)$ of a horizontal $n$-form $L\in \Lambda^n_h(\pi) = \mathcal{F}(\pi)\cdot\pi_{\infty}^*(\Lambda^n(M))$ vanishes on $\mathcal{E}$, the horizontal cohomology class $L + d_h \Lambda^{n-1}_h(\pi)$ produces a unique internal Lagrangian of $\mathcal{E}$ in the following way. According to Noether's identity (to integration by parts), there is a differential form (presymplectic potential current) $\omega_L\in \mathcal{C}\Lambda^1(\mathcal{\pi})\wedge \pi^*_{\infty}(\Lambda^{n-1}(M))$ such that for any $\chi\in \varkappa(\pi)$,
\begin{align}
\mathcal{L}_{E_\chi}(L) = \langle \mathrm{E}(L), \chi \rangle + d_h (E_\chi \lrcorner\, \omega_L)\,.
\label{Noethiden}
\end{align}
Then the differential $n$-form $l = (L + \omega_L)|_{\mathcal{E}}$ represents the corresponding internal Lagrangian, while $dl = d_v\hspace{0.2ex} \omega_L|_{\mathcal{E}}$ represents the presymplectic structure.

If an infinitely prolonged system $\mathcal{E}\subset J^{\infty}(\pi)$ is regular, then each internal Lagrangian is produced by a horizontal $n$-form $L$ such that $\mathrm{E}(L)|_{\mathcal{E}} = 0$ (see Ref.~\cite{DRUZHKOV2023104848}, Theorem 1, and Ref.~\cite{10.1063/1.4828666}, Theorem~3).

\section{The invariant reduction framework \label{InvRedmain}}

Let $\mathcal{E}\subset J^{\infty}(\pi)$ be the infinite prolongation of a regular system of differential equations $F = 0$.
Suppose $X$ is a symmetry of $\mathcal{E}$ having the form $E_{\varphi}|_{\mathcal{E}}$ for some $\varphi\in \varkappa(\pi)$. Let $\mathcal{E}_{X}\subset\mathcal{E}$ be the infinite prolongation of the system for $X$-invariant solutions
\begin{align*}
F = 0\,,\qquad\  \varphi = 0\,.
\end{align*}
Suppose a differential form $\omega\in E_0^{\hspace{0.1ex}p, \hspace{0.2ex} q}(\mathcal{E})$ represents an $X$-invariant element $\Omega$ of a group $E_1^{\hspace{0.1ex}p, \hspace{0.2ex} q}(\mathcal{E})$ of the Vinogradov $\mathcal{C}$-spectral sequence. Then $\mathcal{L}_X \Omega$ is the trivial element of $E_1^{\hspace{0.1ex}p, \hspace{0.2ex} q}(\mathcal{E})$, and the Lie derivative $\mathcal{L}_X \omega$ represents the trivial element of $E_1^{\hspace{0.1ex}p, \hspace{0.2ex} q}(\mathcal{E})$, i.e., there exists $\vartheta\in E_0^{\hspace{0.1ex}p, \hspace{0.2ex} q-1}(\mathcal{E})$ such that
\begin{align}
\mathcal{L}_X \omega = d_0 \vartheta.
\label{mainformula}
\end{align}
Note that $X$ vanishes on $\mathcal{E}_X$. Accordingly, the differential form $\vartheta|_{\mathcal{E}_X}\in E_0^{\hspace{0.1ex}p, \hspace{0.2ex} q-1}(\mathcal{E}_X)$ represents an element $\Theta$ (possibly trivial) of the group $E_1^{\hspace{0.1ex}p, \hspace{0.2ex} q-1}(\mathcal{E}_X)$. If $E_1^{\hspace{0.1ex}p, \hspace{0.2ex} q-1}(\mathcal{E})|_{\mathcal{E}_X} = 0$, then $\Theta$ is unambiguously defined\footnote{The condition $E_1^{\hspace{0.1ex}p, \hspace{0.2ex} q-1}(\mathcal{E})|_{\mathcal{E}_X} = 0$ is equivalent to the inclusion $(\ker d_0^{\hspace{0.1ex}p,\hspace{0.15ex} q-1})|_{\mathcal{E}_X} \subset \mathrm{im}\, d_0^{\hspace{0.1ex}p,\hspace{0.15ex} q-2}$ on $\mathcal{E}_X$.}. In particular, it does not depend on the choice of a representative $\omega\in \Omega$ because $\mathcal{L}_X$ commutes with $d_0$. In this case, the reduction mechanism based on formula~\eqref{mainformula} defines the homomorphism\footnote{If $(p;q)\neq (0;1)$, then $\mathcal{R}_X^{\hspace{0.1ex}p,\hspace{0.2ex} q}$ is well-defined precisely when $E_1^{\hspace{0.1ex}p, \hspace{0.2ex} q-1}(\mathcal{E})|_{\mathcal{E}_X} = 0$.} $\mathcal{R}_X^{\hspace{0.1ex}p,\hspace{0.2ex} q}$ from the vector space of $X$-invariant elements of $E_1^{\hspace{0.1ex}p, \hspace{0.2ex} q}(\mathcal{E})$ to $E_1^{\hspace{0.1ex}p, \hspace{0.2ex} q-1}(\mathcal{E}_X)$, $\mathcal{R}_X^{\hspace{0.1ex}p,\hspace{0.2ex} q}(\Omega) = \Theta$. We also say that $\mathcal{R}^{\hspace{0.1ex} 0,\hspace{0.1ex} 1}_X$ is well-defined if $E^{\hspace{0.1ex} 0, \hspace{0.2ex} 0}_1(\mathcal{E})|_{\mathcal{E}_X} \subset H_{dR}^0(\mathcal{E}_X)$, as we consider reductions of elements of the group $E^{\hspace{0.1ex} 0, \hspace{0.2ex} 1}_1(\mathcal{E})$ modulo additive locally constant functions on $\mathcal{E}_X$. In other words, the invariant reduction of $X$-invariant elements of $E^{\hspace{0.1ex} 0, \hspace{0.2ex} 1}_1(\mathcal{E})$ yields elements\footnote{This situation is similar to the reduction of invariant elements in other groups; recall that, for simplicity, we restrict ourselves to $\mathcal{E}_X$ such that $H^i_{dR}(\mathcal{E}_X) = 0$ for $i > 0$.} of $E^{\hspace{0.1ex} 0, \hspace{0.2ex} 0}_1(\mathcal{E}_X)/H^0_{dR}(\mathcal{E}_X)$. This is the case described in Ref.~\cite{InvRedI} for $n=2$.

\vspace{1ex}

\remarka{As with any reduction, using the inclusion $\mathcal{E}_X\subset \mathcal{E}$, one can restrict \emph{any} elements of $E_r^{\hspace{0.1ex}p, \hspace{0.2ex} q}(\mathcal{E})$ to the system $\mathcal{E}_X$ and get (possibly trivial) elements of $E_r^{\hspace{0.1ex}p, \hspace{0.2ex} q}(\mathcal{E}_X)$. Note also that $E_r^{\hspace{0.1ex}p, \hspace{0.2ex} q}(\mathcal{E}_X)$ can be restricted to subsystems of~$\mathcal{E}_X$.}



\vspace{1ex}

\remarka{This reduction mechanism belongs to the intrinsic geometry of PDEs, viewed as bundles for convenience, i.e., it is not based on inclusions of a differential equation into jet spaces, provided it is a manifold. It also does not require the regularity\footnote{The consideration of the non-regular case is motivated by multi-reduction.} of $\mathcal{E}\subset J^{\infty}(\pi)$ in the general case (see Appendix~\ref{App:A}). The main formula~\eqref{mainformula} relates objects on $\mathcal{E}$ (not on $J^{\infty}(\pi)$ or $\mathcal{E}_X$). From a theoretical point of view, an analysis of the non-triviality of reductions can be performed using a compatibility complex for the linearization operator $l_{\mathcal{E}_X}$ (see Ref.~\cite{krasil1998homological}, Corollary 7.4).
}

\vspace{1ex}



\remarka{\label{RemlMax} For an $\ell$-normal system $\mathcal{E}\subset J^{\infty}(\pi)$ (i.e., a regular manifold such that the equation $\nabla \circ l_{\mathcal{E}} = 0$ for a total differential operator $\nabla\colon P(\mathcal{E})\to \mathcal{F}(\mathcal{E})$ has no nonzero solutions), the groups $E_1^{\hspace{0.1ex}p, \hspace{0.2ex} q}(\mathcal{E})$ are trivial for $q \leqslant n-2$, $(p; q)\neq (0; 0)$ (see Ref.~\cite{VinKr}). In particular, for a system in an extended Kovalevskaya form, the invariant reduction of elements of groups $E_1^{\hspace{0.1ex}p, \hspace{0.2ex} n-1}$ is well-defined (specifically, when $n=2$ and $p = 0$, the result is defined up to an additive locally constant function). In the general case, the situation is more complicated. For example, the group $E^{\hspace{0.1ex} 0, \hspace{0.2ex} n-2}_1$ of the vacuum Maxwell system $d\ast dA = 0$ on a (pseudo-)Riemannian manifold $M^n$ (for $A\in\Lambda^1(M^n)$, $n\geqslant 2$) is non-trivial~\cite{henneaux1997characteristic} (see also Ref.~\cite{krasil1998homological}). The invariant reduction of conservation laws of this system is defined up to the restrictions of elements of its group $E^{\hspace{0.1ex} 0, \hspace{0.2ex} n-2}_1$.}

\vspace{1ex}

The following theorem establishes how the invariant reduction relates the differentials $d_1$ for systems $\mathcal{E}$ and $\mathcal{E}_X$.
In particular, it shows that the invariant reduction relates the cohomology of the invariant subcomplex of $E_1^{\hspace{0.1ex} \bullet, \hspace{0.2ex} q}(\mathcal{E})$ to the cohomology of the complex $E_1^{\hspace{0.1ex}\bullet, \hspace{0.2ex} q-1}(\mathcal{E}_X)$ if $E_1^{\hspace{0.1ex}\bullet, \hspace{0.2ex} q-1}(\mathcal{E})|_{\mathcal{E}_X} = 0$.

\vspace{1ex}

\theorema{\label{Theor1} Let $\mathcal{E}$ be an infinitely prolonged system of differential equations, and let $X = E_{\varphi}|_{\mathcal{E}}$ be its symmetry.
Suppose that the invariant reduction is well-defined for $X$-invariant elements of $E_1^{\hspace{0.1ex}p, \hspace{0.2ex} q}(\mathcal{E})$ and $E_1^{\hspace{0.1ex}p+1, \hspace{0.2ex} q}(\mathcal{E})$. Then on the $X$-invariant subspace of~$E_1^{\hspace{0.1ex}p, \hspace{0.2ex} q}(\mathcal{E})$,}
$$
\mathcal{R}_X^{\hspace{0.1ex}p+1,\hspace{0.2ex} q}\circ d_1 = - d_1\circ\mathcal{R}_X^{\hspace{0.1ex}p,\hspace{0.2ex} q}
$$

\noindent
\textbf{Proof.} If an element $\Omega\in E_1^{\hspace{0.1ex}p, \hspace{0.2ex} q}(\mathcal{E})$ is $X$-invariant, then $d_1\Omega$ is also $X$-invariant.
Denote by $\omega\in E_0^{\hspace{0.1ex}p, \hspace{0.2ex} q}(\mathcal{E})$ a differential form that represents $\Omega\in E_1^{\hspace{0.1ex}p, \hspace{0.2ex} q}(\mathcal{E})$. Let $\vartheta\in E_0^{\hspace{0.1ex}p, \hspace{0.2ex} q-1}(\mathcal{E})$ be a form from~\eqref{mainformula}.
Since $d_v$ commutes with the Lie derivative $\mathcal{L}_X$, we obtain
$$
\mathcal{L}_X d_v\hspace{0.1ex} \omega = d_v \mathcal{L}_X \omega = d_v d_0 \vartheta = d_0 (-d_v\vartheta)\,.
$$
Recall that $d_v\hspace{0.1ex} \omega$ represents $d_1 \Omega$.
Hence, the differential form $-d_v\vartheta|_{\mathcal{E}_X}\in E_0^{\hspace{0.1ex}p+1, \hspace{0.2ex} q-1}(\mathcal{E}_X)$ represents $\mathcal{R}_X^{\hspace{0.1ex}p+1,\hspace{0.2ex} q}(d_1\Omega)$. It also represents $-d_1\mathcal{R}_X^{\hspace{0.1ex}p,\hspace{0.2ex} q}(\Omega)$ because $\vartheta|_{\mathcal{E}_X}\in\mathcal{R}_X^{\hspace{0.1ex}p,\hspace{0.2ex} q}(\Omega)$.

\vspace{1ex}

\remarka{Theorem~\ref{Theor1} remains valid even if no regularity of $\mathcal{E}$ or $\mathcal{E}_X$ is assumed (with the version of the reduction mechanism described in Appendix~\ref{App:A}), and $X$ is an arbitrary symmetry of $\mathcal{E}$ that arises as the restriction of a symmetry of $J^{\infty}(\pi)$ and restricts to $\mathcal{E}_X$.
}

\vspace{1ex}

The interior product of $X$-invariant symmetries and elements of $E_1^{\hspace{0.1ex}p, \hspace{0.2ex} q}(\mathcal{E})$ for $p > 0$ is inherited by $\mathcal{E}_X$ in the following sense.

\vspace{1ex}

\theorema{\label{Theor2} Suppose that $X = E_{\varphi}|_{\mathcal{E}}$, $X_1 = E_{\varphi_1}|_{\mathcal{E}}$ are commuting symmetries of an infinitely prolonged system $\mathcal{E}$. If the invariant reduction is well-defined for $X$-invariant elements of $E_1^{\hspace{0.1ex}p, \hspace{0.2ex} q}(\mathcal{E})$ and $E_1^{\hspace{0.1ex}p-1, \hspace{0.2ex} q}(\mathcal{E})$, then on the $X$-invariant subspace of\,\footnote{If $(p; q) = (1; 1)$, the relation~\eqref{Theor2relation} is understood up to an additive locally constant function on $\mathcal{E}_X$.} $E_1^{\hspace{0.1ex}p, \hspace{0.2ex} q}(\mathcal{E})$,}
\begin{align}\label{Theor2relation}
\mathcal{R}_X^{\hspace{0.1ex}p-1, \hspace{0.2ex} q}\circ {X_1} \lrcorner = -X_1|_{\mathcal{E}_X} \lrcorner\, \circ \mathcal{R}_X^{\hspace{0.1ex}p, \hspace{0.2ex} q}
\end{align}

\noindent
\textbf{Proof.} Suppose a differential form $\omega\in E_0^{\hspace{0.1ex}p, \hspace{0.2ex} q}(\mathcal{E})$ represents an $X$-invariant element of $E_1^{\hspace{0.1ex}p, \hspace{0.2ex} q}(\mathcal{E})$. Let $\vartheta\in E_0^{\hspace{0.1ex}p, \hspace{0.2ex} q-1}(\mathcal{E})$ be a differential form from~\eqref{mainformula}. Since $X_1$ is a symmetry, $\mathcal{L}_{X_1}(E_0^{\hspace{0.1ex}p, \hspace{0.2ex} q}(\mathcal{E}))\subset E_0^{\hspace{0.1ex}p, \hspace{0.2ex} q}(\mathcal{E})$. From a bidegree analysis it follows that $X_1 \lrcorner\,\circ d_0 + d_0\circ X_1 \lrcorner\, = 0$.
Then
$$
\mathcal{L}_X (X_1 \lrcorner\, \omega) = X_1 \lrcorner\, \mathcal{L}_X \omega = X_1 \lrcorner\, d_0 \vartheta = d_0 (-X_1 \lrcorner\, \vartheta)\,,
$$
and the reduction of the element of $E_1^{\hspace{0.1ex}p-1, \hspace{0.2ex} q}(\mathcal{E})$ represented by $X_1 \lrcorner\, \omega$ leads to the element of $E_1^{\hspace{0.1ex}p-1, \hspace{0.2ex} q-1}(\mathcal{E}_X)$ represented by $(-X_1 \lrcorner\, \vartheta)|_{\mathcal{E}_X} = -X_1|_{\mathcal{E}_X} \lrcorner\, \vartheta|_{\mathcal{E}_X}$.

\vspace{1ex}

\remarka{Theorem~\ref{Theor2} remains valid even if no regularity of $\mathcal{E}$ or $\mathcal{E}_X$ is assumed (with the version of the reduction mechanism described in Appendix~\ref{App:A}), and $X$ and $X_1$ are arbitrary symmetries of $\mathcal{E}$ such that they arise as restrictions of symmetries of $J^{\infty}(\pi)$, restrict to $\mathcal{E}_X$, and their commutator $[X, X_1]$ is a trivial symmetry.
}

\subsection{Multi-reduction}\label{Sectionmultired}

In contrast to the multi-reduction method proposed in Ref.~\cite{anderson1997symmetry}, the approach based on formula~\eqref{mainformula} requires a one-dimensional symmetry algebra.
The idea to apply the reduction mechanism step by step using solvable subalgebras of Lie algebras of symmetries leads to the following observation. Suppose that $X_1 = E_{\varphi_1}|_{\mathcal{E}}$ is another symmetry of $\mathcal{E}$ and $[X, X_1] = cX$ for some $c\in \mathbb{R}$. Then the system $\mathcal{E}_X$ inherits the symmetry $X_1|_{\mathcal{E}_X}$. Indeed,
\begin{align*}
E_{c\varphi}|_{\mathcal{E}} = c E_{\varphi}|_{\mathcal{E}} = [X, X_1] = [E_{\varphi}, E_{\varphi_1}]|_{\mathcal{E}} = E_{\{\varphi, \varphi_1\}}|_{\mathcal{E}}\,,\qquad \{\varphi, \varphi_1\} = E_{\varphi}(\varphi_1) - E_{\varphi_1}(\varphi)\,,
\end{align*}
and hence, $c \varphi|_{\mathcal{E}} = E_{\varphi}(\varphi_1)|_{\mathcal{E}} - E_{\varphi_1}(\varphi)|_{\mathcal{E}}$. Restricting to $\mathcal{E}_X$, we see that $E_{\varphi_1}(\varphi)|_{\mathcal{E}_X} = 0$, i.e., $E_{\varphi_1}$ is tangent to $\mathcal{E}_X$. In other words, $X_1|_{\mathcal{E}_X}$ is a symmetry of $\mathcal{E}_X$.

\vspace{1ex}

\remarka{This reasoning can be naturally generalized to the case of arbitrary solvable algebras of symmetries and is undoubtedly known. We have presented it here to introduce the notation.}

\vspace{1ex}

If $X_1|_{\mathcal{E}_X}$ is non-trivial, one can use it to perform an additional step of the reduction for $X$-invariant elements of a group $E_1^{\hspace{0.1ex}p, \hspace{0.2ex} q}(\mathcal{E})$, $2 \leqslant q \leqslant n$, provided that the first step gives a non-trivial $X_1|_{\mathcal{E}_X}$-invariant element of the group $E_1^{\hspace{0.1ex}p, \hspace{0.2ex} q-1}(\mathcal{E}_X)$. However, these assumptions are inconsistent for $X_1$-invariant elements of $E_1^{\hspace{0.1ex}p, \hspace{0.2ex} q}(\mathcal{E})$ if $X_1$ is not $X$-invariant (i.e., if $X$ and $X_1$ do not commute).

\vspace{1ex}

\propositiona{\label{Proposition} Let $X = E_{\varphi}|_{\mathcal{E}}$ and $X_1 = E_{\varphi_1}|_{\mathcal{E}}$ be symmetries of an infinitely prolonged system $\mathcal{E}$ such that $[X, X_1] = cX$ for some $c\in \mathbb{R}$. Suppose $\omega\in E_0^{\hspace{0.1ex}p, \hspace{0.2ex} q}(\mathcal{E})$ represents an element of $E_1^{\hspace{0.1ex}p, \hspace{0.2ex} q}(\mathcal{E})$ $(\text{here}\ 2 \leqslant q\leqslant n)$ that is both $X$-invariant and $X_1$-invariant, and $\mathcal{R}^{\hspace{0.1ex}p, \hspace{0.2ex} q}_X$ is well-defined. Let $\mathcal{L}_X \omega = d_0 \vartheta$.\\
1) If $\vartheta|_{\mathcal{E}_X}$ represents an $X_1|_{\mathcal{E}_X}$-invariant non-trivial element of $E_1^{\hspace{0.1ex}p, \hspace{0.2ex} q-1}(\mathcal{E}_X)$, then $c = 0$.\\
2) If $c = 0$, then $\vartheta|_{\mathcal{E}_X}$ represents an $X_1|_{\mathcal{E}_X}$-invariant element of $E_1^{\hspace{0.1ex}p, \hspace{0.2ex} q-1}(\mathcal{E}_X)$.}

\vspace{0.5ex}
\noindent
\textbf{Proof.}
There is $\vartheta_1\in E_0^{\hspace{0.1ex}p, \hspace{0.2ex} q-1}(\mathcal{E})$ such that $\mathcal{L}_{X_1}\omega = d_0\vartheta_1$ and hence,
$$
d_0(c\vartheta) = \mathcal{L}_{cX}\omega = \mathcal{L}_{[X,\, X_1]}\hspace{0.2ex}\omega = [\mathcal{L}_{X}, \mathcal{L}_{X_1}]\hspace{0.2ex}\omega = \mathcal{L}_{X} d_0 \vartheta_1 - \mathcal{L}_{X_1} d_0 \vartheta = d_0 (\mathcal{L}_{X}\vartheta_1 - \mathcal{L}_{X_1}\vartheta)\,.
$$
Then we get
\begin{align}
c\vartheta - (\mathcal{L}_{X}\vartheta_1 - \mathcal{L}_{X_1}\vartheta)\in \ker d_0\,.
\label{reducrelat}
\end{align}
Here $(\ker d_0^{\hspace{0.1ex} p, \hspace{0.2ex} q-1})|_{\mathcal{E}_X} \subset \mathrm{im}\, d_0^{\hspace{0.1ex} p, \hspace{0.2ex} q-2}$. Then for the restrictions to $\mathcal{E}_X$, one has
\begin{align}
c\vartheta|_{\mathcal{E}_X} + \mathcal{L}_{X_1|_{\mathcal{E}_X}}(\vartheta|_{\mathcal{E}_X}) \in \mathrm{im}\, d_0\,,
\label{multireduction}
\end{align}
where $d_0$ denotes the differential on $\mathcal{E}_X$.\\
1) If $\vartheta|_{\mathcal{E}_X}$ determines an $X_1|_{\mathcal{E}_X}$-invariant element of $E_1^{\hspace{0.1ex}p, \hspace{0.2ex} q-1}(\mathcal{E}_X)$, then $\mathcal{L}_{X_1|_{\mathcal{E}_X}}(\vartheta|_{\mathcal{E}_X}) \in \mathrm{im}\, d_0$. In this case, $c\vartheta|_{\mathcal{E}_X} \in \mathrm{im}\, d_0$ and $\vartheta|_{\mathcal{E}_X}$ cannot represent a non-trivial element of $E_1^{\hspace{0.1ex}p, \hspace{0.2ex} q-1}(\mathcal{E}_X)$ whenever $c\neq 0$.\\
2) If $c = 0$ in~\eqref{multireduction}, then $\vartheta|_{\mathcal{E}_X}$ represents an $X_1|_{\mathcal{E}_X}$-invariant element of $E_1^{\hspace{0.1ex}p, \hspace{0.2ex} q-1}(\mathcal{E}_X)$.

\vspace{1ex}

\remarka{From~\eqref{reducrelat} it follows that when $c = 0$ and $E_1^{\hspace{0.1ex}p, \hspace{0.2ex} q-1}(\mathcal{E}) = 0$, the reduction of an element of $E_1^{\hspace{0.1ex}p, \hspace{0.2ex} q}(\mathcal{E})$ first under $X$ and then under $X_1$ differs by a sign from its reduction first under $X_1$ and then under $X$, provided that all reductions are well-defined.}

\vspace{1ex}

\remarka{Proposition~\ref{Proposition} remains valid even if no regularity of $\mathcal{E}$ or $\mathcal{E}_X$ is assumed (with the version of the reduction mechanism described in Appendix~\ref{App:A}), and $X$ and $X_1$ are arbitrary symmetries of $\mathcal{E}$ such that they arise as restrictions of symmetries of $J^{\infty}(\pi)$, restrict to $\mathcal{E}_X$, and their commutator $[X, X_1]$ is equivalent to $cX$.
}


\subsection{\label{NoetherTheo} An analog of Noether's theorem for invariant solutions}

Let us recall how the Noether theorem formulated in a slightly generalized form relates certain symmetries and conservation laws of regular systems. Suppose $\Omega\in E_1^{\hspace{0.1ex}2, \hspace{0.2ex} n-1}(\mathcal{E})$ is a presymplectic structure of an infinitely prolonged system $\mathcal{E}$. If $X_1 = E_{\varphi_1}|_{\mathcal{E}}$ is a symmetry of $\mathcal{E}$, then the variational $1$-form $X_1 \lrcorner\, \Omega$ is well-defined. Assume that $X_1$ is a \emph{Noether symmetry} for $\Omega$, i.e., $X_1 \lrcorner\, \Omega$ is $d_1$-exact. Then $X_1$ corresponds to the conservation laws whose differentials $d_1$ coincide with $X_1 \lrcorner\, \Omega$. In other words, a Noether symmetry $X_1$ corresponds to a conservation law $\xi\in E^{\hspace{0.1ex}0, \hspace{0.2ex}n-1}_1(\mathcal{E})$ if
\begin{align}
X_1\lrcorner\, \Omega = d_1\xi\,.
\label{Noethercorr}
\end{align}

\remarka{If $\Omega$ is generated by an internal Lagrangian of $\mathcal{E}$, the same correspondence, for certain symmetries $X_1$, follows from Noether's theorem for action functionals on jet spaces (see Ref.~\cite{Olver} for Lagrangian systems, Theorem~5.58). Namely, an internal Lagrangian generating $\Omega$ is produced by a horizontal form $L\in \Lambda^n_h(\pi)$ such that the variational derivative $\mathrm{E}(L)\in \widehat{\varkappa}(\pi)$ vanishes on $\mathcal{E}$ (see Ref.~\cite{DRUZHKOV2023104848}, Theorem~1). If an evolutionary field $E_{\varphi_1}$ is a variational symmetry~\cite{Olver} of $\mathrm{E}(L) = 0$ (that is, it preserves the horizontal cohomology class of $L$), it follows from Noether's identity~\eqref{Noethiden} that $\langle \mathrm{E}(L), \varphi_1 \rangle$ is $d_h$-exact. The restriction of a potential for $\langle \mathrm{E}(L), \varphi_1 \rangle$ to $\mathcal{E}$ represents a conservation law corresponding to $X_1 = E_{\varphi_1}|_{\mathcal{E}}$ in the sense of~\eqref{Noethercorr}, provided that $E_{\varphi_1}$ restricts to $\mathcal{E}$. However, this requirement is more limiting than~\eqref{Noethercorr}, which may apply to symmetries of $\mathcal{E}$ that do not extend to symmetries of $\mathrm{E}(L) = 0$. Note also that $X_1\mapsto X_1\lrcorner\, \Omega$ is a correspondence between arbitrary symmetries and variational $1$-forms of $\mathcal{E}$. It is surjective if $\Omega$ is produced by $L$ such that the infinite prolongation of $\mathrm{E}(L) = 0$ coincides with $\mathcal{E}$.}

\vspace{1ex}

\remarka{If $X$ is a Noether symmetry, then $\Omega$ is $X$-invariant, because $\mathcal{L}_X \Omega = X\lrcorner \,(d_1\Omega) + d_1 (X\lrcorner \,\Omega) = 0$, which follows from $d_1\Omega = 0$ and the identity $d_1\circ d_1 = 0$. If $\mathcal{E}$ is $\ell$-normal, then the group $E_2^{\hspace{0.1ex}1, \hspace{0.2ex} n-1}(\mathcal{E})$ is trivial (see, e.g, Ref.~\cite{VinKr}), and therefore, the $X$-invariance of $\Omega$ implies that $X$ is a Noether symmetry. In addition, in the $\ell$-normal case, the group $E_2^{\hspace{0.1ex}0, \hspace{0.2ex} n-1}(\mathcal{E})$ is also trivial (e.g., Ref.~\cite{VinKr}), and each Noether symmetry corresponds to a unique conservation law.}

\vspace{1ex}

If $p = 2$, $q = n-1$, from~\eqref{mainformula} we see that $\vartheta|_{\mathcal{E}_X}$ represents an element of $E_1^{\hspace{0.1ex}2, \hspace{0.2ex} n-2}(\mathcal{E}_X)$. \emph{If $Y$ is a symmetry of $\mathcal{E}_X$, and the element of $E_1^{\hspace{0.1ex}1, \hspace{0.2ex} n-2}(\mathcal{E}_X)$ represented by $Y \lrcorner\, (\vartheta|_{\mathcal{E}_X})$ is $d_1$-exact, then $Y$ gives rise to an element of $E_1^{\hspace{0.1ex}0, \hspace{0.2ex} n-2}(\mathcal{E}_X)$.} This result can be considered Noether's theorem for invariant solutions.
A particuarly interesting case here is $n = 2$, because the resulting element of $E_1^{\hspace{0.1ex}0, \hspace{0.2ex} n-2}(\mathcal{E}_X)$ is a constant of $X$-invariant motion. In this case, the $d_1$-exactness of the element of $E_1^{\hspace{0.1ex}1, \hspace{0.2ex} n-2}(\mathcal{E}_X)$ given by $Y \lrcorner\, (\vartheta|_{\mathcal{E}_X})$ means that the differential form $Y \lrcorner\, (\vartheta|_{\mathcal{E}_X})$ is $d$-exact.
One can establish the exactness of a differential $1$-form and find the corresponding potential using standard methods from finite-dimensional differential geometry.

\vspace{1ex}

\remarka{One can contract two symmetries of $\mathcal{E}_X$ with a reduction of an $X$-invariant presymplectic structure of $\mathcal{E}$ to obtain an element of $E_1^{\hspace{0.1ex}0, \hspace{0.2ex} n-2}(\mathcal{E}_X)$.}

\vspace{1ex}

The invariant system $\mathcal{E}_X$ inherits the Noether correspondence defined by~\eqref{Noethercorr}. This connection between the Noether theorem and its version for invariant solutions is given by the following theorem.

\vspace{1ex}

\theorema{\label{Theor3} Suppose $X = E_{\varphi}|_{\mathcal{E}}$, $X_1 = E_{\varphi_1}|_{\mathcal{E}}$ are commuting symmetries of an $\ell$-normal system of equations $\mathcal{E}$. Let $\Omega\in E_1^{\hspace{0.1ex}2, \hspace{0.2ex} n-1}(\mathcal{E})$ be an $X$-invariant presymplectic structure, and let $\xi\in E_1^{\hspace{0.1ex}0, \hspace{0.2ex} n-1}(\mathcal{E})$ be a conservation law of $\mathcal{E}$ such that $X_1\lrcorner\, \Omega = d_1\xi$. Then}
$$
X_1|_{\mathcal{E}_X} \lrcorner\, \mathcal{R}_X^{\hspace{0.1ex}2, \hspace{0.2ex} n-1}(\Omega) = d_1 \mathcal{R}_X^{\hspace{0.1ex}0, \hspace{0.2ex} n-1}(\xi)
$$

\noindent
\textbf{Proof.} Since $\mathcal{E}$ is $\ell$-normal, $E_2^{\hspace{0.1ex}0, \hspace{0.2ex} n-1}(\mathcal{E}) = 0$. It follows that the relation
$$
d_1 \mathcal{L}_X \xi = \mathcal{L}_X d_1\xi = \mathcal{L}_X (X_1\lrcorner\, \Omega) = X_1\lrcorner\, \mathcal{L}_X \Omega = 0
$$
implies $\mathcal{L}_X \xi = 0$, i.e., the conservation law $\xi$ is $X$-invariant, as well as $X_1\lrcorner\, \Omega$. Remark~\ref{RemlMax} shows that the reductions of $\xi$, $X_1\lrcorner\, \Omega$, and $\Omega$ are well-defined. Applying the mapping $\mathcal{R}_X^{\hspace{0.1ex}1, \hspace{0.2ex} n-1}$ to the relation
$X_1 \lrcorner\, \Omega = d_1 \xi$ and using Theorem~\ref{Theor1} and Theorem~\ref{Theor2}, we find
$$
-X_1|_{\mathcal{E}_X} \lrcorner\, \mathcal{R}_X^{\hspace{0.1ex}2, \hspace{0.2ex} n-1}(\Omega) = -d_1 \mathcal{R}_X^{\hspace{0.1ex}0, \hspace{0.2ex} n-1}(\xi)\,.
$$

\subsection{Reduction of variational principles}

The reduction mechanism is homological and can be applied to $X$-invariant elements of groups $\widetilde{E}_1^{\hspace{0.1ex} 0, \hspace{0.2ex} k}(\mathcal{E})$ given by~\eqref{canonisom}. If $\omega\in \widetilde{E}_0^{\hspace{0.1ex} 0, \hspace{0.2ex} k}(\mathcal{E})$ represents an $X$-invariant element of $\widetilde{E}_1^{\hspace{0.1ex} 0, \hspace{0.2ex} k}(\mathcal{E})$, there is $\vartheta\in \widetilde{E}_0^{\hspace{0.1ex} 0, \hspace{0.2ex} k-1}(\mathcal{E})$ such that
$$
\mathcal{L}_X \omega = \tilde{d}_0 \vartheta.
$$
In particular, this yields the reduction mechanism for $X$-invariant internal Lagrangians of an infinitely prolonged system $\mathcal{E}$. If $\widetilde{E}^{\hspace{0.1ex} 0, \hspace{0.2ex} n-2}_1(\mathcal{E})|_{\mathcal{E}_X} = 0$, the reduction is well-defined. The spectral sequence for Lagrangian formalism~\cite{DRUZHKOV2023104848} shows that $\ell$-normal systems are of this type. Theorem~\ref{Theor1} also admits a straightforward generalization to this case (see Appendix~\ref{App:B}), as does Proposition~\ref{Proposition}.

\vspace{1ex}

\remarka{The Noether theorem relates symmetries of internal Lagrangians of $\mathcal{E}$ and its conservation laws~\cite{DRUZHKOV2024105143}. Similarly, the Noether theorem for PDE systems satisfied by invariant solutions can be formulated as a relation between symmetries of a reduction of an $X$-invariant internal Lagrangian and elements of the group $E^{\hspace{0.1ex} 0, \hspace{0.2ex} n-2}_1(\mathcal{E}_X)$.}

\vspace{1ex}

Internal Lagrangians of PDEs can be interpreted as variational principles~\cite{DRUZHKOV2024105143, DRUZHKOV2024ATMP}.
The invariant reduction of an internal Lagrangian leads to an element of the group $\widetilde{E}_1^{\hspace{0.1ex} 0, \hspace{0.2ex} n-2}(\mathcal{E}_X)$. Such elements determine variational principles in a similar way. For the sake of simplicity, we restrict ourselves to variational principles determined by elements of the group $\widetilde{E}_1^{\hspace{0.15ex} 0, \hspace{0.2ex} 0}(\mathcal{E}_X)$. In terms of the reduction under a single symmetry, they appear if $n=2$.

Suppose a differential form $\varrho \in \Lambda^1(\mathcal{E}_X)$ represents an element of $\widetilde{E}_1^{\hspace{0.1ex} 0, \hspace{0.2ex} 0}(\mathcal{E}_X)$, i.e., $d\varrho \in \mathcal{C}^2\Lambda^2(\mathcal{E}_X)$. Denote by $\pi_{\mathcal{E}_X}$ the projection $\pi_{\mathcal{E}}|_{\mathcal{E}_X}$. Let $\gamma\colon \mathbb{R}\times M\to \mathcal{E}_X$ be a mapping such that all $\gamma(\tau)\colon M\to \mathcal{E}_X$, $\gamma(\tau)\colon x\mapsto \gamma(\tau, x)$ are sections of~$\pi_{\mathcal{E}_X}$ (for $\tau\in \mathbb{R}$). Then $\gamma(\tau)$ is a path in sections of~$\pi_{\mathcal{E}_X}$.

\vspace{1ex}

\definitiona{\label{Def1} A section $\sigma\colon M\to \mathcal{E}_X$ of $\pi_{\mathcal{E}_X}$ is a \emph{stationary point} of $\varrho + d(\mathcal{F}(\mathcal{E}_X))\in \widetilde{E}^{\hspace{0.15ex}0, \hspace{0.2ex} 0}_1(\mathcal{E}_X)$, if the relation
\begin{align}
\dfrac{d}{d\tau}\Big|_{\tau = 0}\int_N \gamma(\tau)^*(\varrho) = 0
\label{varpr}
\end{align}
holds for any embedded, compact, $1$-dimensional submanifold $N\subset M$ and any path $\gamma(\tau)$ in sections of $\pi_{\mathcal{E}_X}$ such that $\gamma(0) = \sigma$ and each point of the boundary $\partial N$ is fixed.}

\vspace{1ex}
\noindent
We assume that each appropriate $N$ is oriented. The choice of a representative plays no role since the boundary $\partial N$ remains fixed (or empty).

\vspace{1ex}

\propositiona{All solutions of the system $\pi_{\mathcal{E}_X}$ are stationary points of any element of $\widetilde{E}_1^{\hspace{0.1ex} 0, \hspace{0.2ex} 0}(\mathcal{E}_X)$.}

\vspace{0.5ex}
\noindent
\textbf{Proof.} Denote by $0_M$ the zero section $0_M\colon M \to \mathbb{R}\times M$, $0_M(x) = (0, x)$. The former summand on the RHS of the homotopy formula
\begin{align}
\dfrac{d}{d\tau}\Big|_{\tau = 0}\int_N \gamma(\tau)^*(\varrho) = \int_N d\, 0_M^*(\partial_{\tau} \lrcorner\, \gamma^*(\varrho)) + \int_N  0_M^*(\partial_{\tau} \lrcorner\, \gamma^*(d\varrho))
\label{thehomotform}
\end{align}
vanishes for any section taken as $\gamma(0)$ provided that each point of the boundary $\partial N$ remains fixed. The latter summand vanishes if $\gamma(0)$ is a solution of $\pi_{\mathcal{E}_X}$ since $d\varrho\in \mathcal{C}^2\Lambda^2(\mathcal{E}_X)$ and $0_M^*\circ \gamma^* = \gamma(0)^*$.

\vspace{1ex}

It follows from the proof that the variational principle is determined by $d\varrho\in E_1^{\hspace{0.1ex} 2,\hspace{0.2ex} 0}(\mathcal{E}_X)$, which defines a field of operators from $\pi_{\mathcal{E}_X}$-vertical tangent vectors to Cartan forms.
As in classical Hamiltonian mechanics, if $\mathcal{E}_X$ is a finite-dimensional smooth manifold and the field of operators is non-degenerate at each point of $\mathcal{E}_X$, the variational principle yields only solutions to $\pi_{\mathcal{E}_X}$. In this case, the restrictions of $d\varrho$ (or $-d\varrho$) to fibers of $\pi_{\mathcal{E}_X}$ are invertible and determine a Poisson bivector, which maps differentials of constants of $X$-invariant motion to symmetries of $\mathcal{E}_X$. This follows from the fact that the (local) flow of a vector field corresponding to a constant of $X$-invariant motion preserves the differential form $d\varrho$ (since its interior product with $d\varrho$ is an exact $1$-form). Then it preserves the kernel of $d\varrho$ on $\mathcal{E}_X$, i.e., the Cartan distribution.

\vspace{1ex}

\remarka{For $k\geqslant 0$, elements of groups $\widetilde{E}^{\hspace{0.1ex} 0, \hspace{0.2ex} k}_1(\mathcal{E})$ also determine variational principles (in terms of $k$-dimensional ``instantaneous states''). When $k > 0$, analogs of spatial equations and spatial-gauge equivalence~\cite{DRUZHKOV2024ATMP} may be required. An alternative approach is to deal with analogs of \emph{intrinsic} Lagrangians~\cite{GRIGORIEV2016, GRIGORIEV2022115686}. In the case $k = 0$, these two approaches coincide.
}

\section{Computational algorithms for evolution systems \label{Computal}}

In some cases, the reduction of symmetry-invariant structures can be described algorithmically. Below we provide two reduction algorithms: one for conservation laws of arbitrary systems of evolution equations, and one for presymplectic structures of $(1+1)$-dimensional systems of evolution equations.

Let us consider a system of evolution equations $F = 0$, where $P(\pi) = \varkappa(\pi)$, $t$ denotes $x^1$,
\begin{align*}
F^i = u^i_t - f^i\,,
\end{align*}
and all functions $f^1$, \ldots, $f^m$ do not depend on $u^j_{t + \alpha}$ for $|\alpha|\geqslant 0$, $1\leqslant j \leqslant m$. We treat the variables $u^i_{t + \alpha}$ as external coordinates associated with the corresponding infinite prolongation $\mathcal{E}$. Other coordinates on $J^{\infty}(\pi)$ can be treated as local coordinates on the system $\mathcal{E}$. Then there is the inclusion of the algebras $\mathcal{F}(\mathcal{E})\subset \mathcal{F}(\pi)$, leading to the inclusion of the modules $\varkappa(\mathcal{E})\subset \varkappa(\pi)$.

{Within this section, we consider only systems of evolution equations.}
Suppose that $\varphi\in \varkappa(\mathcal{E})\subset \varkappa(\pi)$ is the characteristic of a symmetry $E_{\varphi}$ of the system $\mathcal{E}$. Then
\begin{align*}
E_{\varphi}(F) = l_{\varphi}(F)\,,
\end{align*}
where $l_{\varphi}(F) = E_F(\varphi)$. Let us recall that the $i^{\rm th}$ component of $E_{\varphi}(F)\in P(\pi)$ is $E_{\varphi}(F^i) = D_{\alpha}(\varphi^j)\, \partial F^i/\partial u^j_{\alpha}$. Denote $X = E_{\varphi}|_{\mathcal{E}}$. Since all evolution systems are $\ell$-normal, the reduction of $X$-invariant elements of $E_1^{\hspace{0.1ex}p, \hspace{0.2ex} n-1}(\mathcal{E})$ is well-defined ($p\geqslant 0$).

\subsection{\label{Redofconslaws} The invariant reduction of conservation laws}

Conservation laws of evolution equations can be described both in terms of their characteristics~\cite{Olver} and in terms of their cosymmetries~\cite{VinKr}. Analogous to the inclusion $\mathcal{F}(\mathcal{E}) \subset \mathcal{F}(\pi)$, elements of $E_0^{\hspace{0.1ex}0, \hspace{0.2ex} n-1}(\mathcal{E})$ can be interpreted as horizontal $(n-1)$-forms on $J^{\infty}(\pi)$. If $\omega\in E_0^{\hspace{0.1ex}0, \hspace{0.2ex} n-1}(\mathcal{E})$ represents a conservation law of an infinitely prolonged system of evolution equations $\mathcal{E}$, then there is a total differential operator $A\colon P(\pi)\to \Lambda^{n}_h(\pi)$ such that
$$
d_h\hspace{0.2ex}\omega = AF\,.
$$
Integrating by parts, we find a unique homomorphism $\psi\in \mathrm{Hom}_{\mathcal{F}(\pi)} (P(\pi), \Lambda_h^n(\pi))$ such that for some total differential operator $A_1\colon P(\pi)\to \Lambda^{n-1}_h(\pi)$, the relation
\begin{align}
d_h\hspace{0.2ex} \omega = AF = \langle \psi , F \rangle + d_h(A_1 F)
\label{CLcharform}
\end{align}
holds on $J^{\infty}(\pi)$. The homomorphism $\psi$ is a characteristic of the conservation law. Note that its components $\psi_1, \ldots, \psi_m$ do not depend on the variables of the form $u^j_{t + \alpha}$ (see, e.g., Ref.~\cite{MartinezAlonso}, Lemma~3). Here
\begin{align}
\langle \psi, F \rangle = \psi_i F^i dt\wedge dx^2\wedge\ldots\wedge dx^{n}\,.
\label{cosymcompon}
\end{align}
Then $\psi$ can be considered the cosymmetry of the conservation law.

The conservation law is $X$-invariant if and only if the cosymmetry $E_{\varphi}(\psi) + l_{\varphi}^{\hspace{0.2ex} *}(\psi)$ of its Lie derivative (represented by $\mathcal{L}_{X} \omega$) is zero. Therefore, in order to construct a potential for $\mathcal{L}_{X} \omega$, one can use the following observation.
\begin{align*}
d_h\hspace{0.2ex} \mathcal{L}_{E_{\varphi}} \omega &= \mathcal{L}_{E_{\varphi}} d_h\hspace{0.2ex} \omega = \mathcal{L}_{E_{\varphi}} \langle \psi , F \rangle + \mathcal{L}_{E_{\varphi}}d_h(A_1F) = \langle E_{\varphi} (\psi), F \rangle + \langle \psi, E_{\varphi}(F) \rangle + d_h \mathcal{L}_{E_{\varphi}}(A_1F) {}\\
&= \langle E_{\varphi} (\psi), F \rangle + \langle \psi, l_{\varphi}(F) \rangle + d_h \mathcal{L}_{E_{\varphi}}(A_1F)\,.
\end{align*}
Integrating by parts, we obtain
\begin{align}
\langle \psi, l_{\varphi}(F) \rangle = \langle l_{\varphi}^{\hspace{0.2ex} *}(\psi), F \rangle + d_h (A_2F)\,,
\label{intparts}
\end{align}
where $A_2\colon P(\pi)\to \Lambda^{n-1}_h(\pi)$ is a total differential operator. If the conservation law is $X$-invariant, we have $E_{\varphi}(\psi) + l_{\varphi}^{\hspace{0.2ex} *}(\psi) = 0$, and hence,
\begin{align*}
d_h\hspace{0.2ex} \mathcal{L}_{E_{\varphi}} \omega = d_h (\mathcal{L}_{E_{\varphi}}(A_1F) + A_2F)\,.
\end{align*}
Assuming that the de Rham cohomology group $H^{n-1}_{dR}(M)$ is trivial, one can apply the total homotopy formula~\cite{Olver} to find a horizontal form $\widehat{\vartheta} \in \Lambda^{n-2}_h(\pi)$ such that on $J^{\infty}(\pi)$,
\begin{align}
\mathcal{L}_{E_{\varphi}} \omega - \mathcal{L}_{E_{\varphi}}(A_1F) - A_2F = d_h \widehat{\vartheta}\,.
\label{algoformula}
\end{align}
Restricting to $\mathcal{E}$, we get $\mathcal{L}_{X} \omega = d_0 \widehat{\vartheta}|_{\mathcal{E}}$\hspace{0.1ex}, since $\varphi$ is a characteristic of a symmetry. The potential $\vartheta = \widehat{\vartheta}|_{\mathcal{E}}\in E^{\hspace{0.1ex}0, \hspace{0.2ex} n-2}_0(\mathcal{E})$ leads to the desired reduction of the conservation law. The reduction algorithm thus consists of the two integrations by parts \eqref{CLcharform}, \eqref{intparts}, and an application of the total homotopy formula.

\subsection{\label{Redofpresstr} The invariant reduction of presymplectic structures for (1+1)-dimensional systems of evolution equations}

Every presymplectic structure of an $\ell$-normal system is generated by an internal Lagrangian~\cite{DRUZHKOV2024105143}. Then all presymplectic structures of (1+1)-dimensional systems of evolution equations originate from the stationary action principle. They can be described in terms of presymplectic operators reproducing the same correspondence between symmetries and variational $1$-forms (see Section~\ref{NoetherTheo}). 

Denote
$$
\widehat{P}(\pi) = \mathrm{Hom}_{\mathcal{F}(\pi)} (P(\pi), \Lambda_h^n(\pi))\,,\qquad
\widehat{P}(\mathcal{E}) = \widehat{P}(\pi)|_{\mathcal{E}}\,.
$$
The relation between presymplectic structures and presymplectic operators 
can be described as follows. Let $L\in \Lambda^n_h(\pi)$ be a horizontal $n$-form such that the variational derivative $\mathrm{E}(L)\in \widehat{\varkappa}(\pi)$ vanishes on $\mathcal{E}$. Then $L$ gives rise to a presymplectic structure of $\mathcal{E}$. Besides, for some total differential operator $B\colon P(\pi)\to \widehat{\varkappa}(\pi)$,
\begin{align}
\mathrm{E}(L) = BF\,.
\label{variop}
\end{align}
The restriction $\Delta\colon \varkappa(\mathcal{E})\to \widehat{P}(\mathcal{E})$ of the formally adjoint operator $B^*\colon \varkappa(\pi)\to \widehat{P}(\pi)$ to $\mathcal{E}$ is a \textit{presymplectic operator} corresponding to the presymplectic structure.
The presymplectic structure is $E_{\varphi}|_{\mathcal{E}}$-invariant if and only if for the cosymmetry $\psi = \Delta(\varphi)$, the presymplectic operator $l_{\psi} - l_{\psi}^{\,*}$ is zero.

\vspace{1ex}

\remarka{A more precise description of elements of $\ker d_1^{\hspace{0.1ex} 2,\,n-1}$ in terms of presymplectic operators is given in Ref.~\cite{VinKr}. According to it, for a total differential operator $\square\colon P(\mathcal{E})\to \widehat{P}(\mathcal{E})$ such that $\square = \square^*$, the operator $\Delta + \square\circ l_{\mathcal{E}}$ is a presymplectic operator corresponding to the same presymplectic structure. Note that each class of presymplectic operators of a system of evolution equations contains a unique operator that does not involve the total derivative $\,\overline{\!D}_t$.} 

\vspace{1ex}

Suppose $\omega\in E^{\hspace{0.1ex}2, \hspace{0.2ex} 1}_0(\mathcal{E})$ represents an $X$-invariant presymplectic structure of an infinitely prolonged (1+1)-dimensional system $\mathcal{E}$ of evolution equations. Its Lie derivative $\mathcal{L}_X \omega$ has the form
\begin{align*}
\mathcal{L}_X\omega = \gamma_{ij1}^{ks}\,\overline{\!\theta}^{\hspace{0.2ex} i}_{kx}\wedge \,\overline{\!\theta}^{\hspace{0.2ex} j}_{sx}\wedge dx + \gamma_{ij2}^{ks}\ \overline{\!\theta}^{\hspace{0.2ex} i}_{kx}\wedge \,\overline{\!\theta}^{\hspace{0.2ex} j}_{sx}\wedge dt
\end{align*}
in local coordinates. Here $\,\overline{\!\theta}^{\hspace{0.2ex} i}_{kx} = \theta^{\hspace{0.2ex} i}_{kx}|_{\mathcal{E}}$.
The corresponding differential form $\vartheta\in E^{\hspace{0.1ex}2, \hspace{0.2ex} 0}_0(\mathcal{E}) = \mathcal{C}^2\Lambda^2(\mathcal{E})$ from~\eqref{mainformula} is unique, as further analysis shows. It satisfies the equation
\begin{align}
\gamma_{ij1}^{ks}\,\overline{\!\theta}^{\hspace{0.2ex} i}_{kx}\wedge \,\overline{\!\theta}^{\hspace{0.2ex} j}_{sx}\wedge dx = dx\wedge\mathcal{L}_{\,\overline{\!D}_x}\vartheta\,,\qquad \,\overline{\!D}_x = D_x|_{\mathcal{E}}\,.
\label{thetaequation}
\end{align}
Integrating by parts in $\gamma_{ij1}^{ks}\,\overline{\!\theta}^{\hspace{0.2ex} i}_{kx}\wedge \,\overline{\!\theta}^{\hspace{0.2ex} j}_{sx} = (\gamma_{ij1}^{ks} - \gamma_{ji1}^{sk})\,\overline{\!\theta}^{\hspace{0.2ex} i}_{kx}\otimes \,\overline{\!\theta}^{\hspace{0.2ex} j}_{sx}$,
\begin{align*}
(\gamma^{ks}_{ij1} - \gamma^{sk}_{ji1})\,\overline{\!\theta}^{\hspace{0.2ex} i}_{kx}\otimes \,\overline{\!\theta}^{\hspace{0.2ex} j}_{sx} = \mathcal{L}_{\,\overline{\!D}_x} \big((\gamma^{ks}_{ij1} - \gamma^{sk}_{ji1})\,\overline{\!\theta}^{\hspace{0.2ex} i}_{kx}\otimes \,\overline{\!\theta}^{\hspace{0.2ex} j}_{(s-1)x}\big) - \mathcal{L}_{\,\overline{\!D}_x} \big((\gamma^{ks}_{ij1} - \gamma^{sk}_{ji1})\,\overline{\!\theta}^{\hspace{0.2ex} i}_{kx}\big)\otimes \,\overline{\!\theta}^{\hspace{0.2ex} j}_{(s-1)x} = \ldots
\end{align*}
and applying the antisymmetrization, we find
\begin{align}
\vartheta &= \dfrac{1}{2}\sum_{s=1}^{+\infty}\sum_{r = 1}^s (-1)^{r-1} \mathcal{L}_{\,\overline{\!D}_x}^{r-1}\big((\gamma^{ks}_{ij1} - \gamma^{sk}_{ji1}) \,\overline{\!\theta}^{\hspace{0.2ex} i}_{kx}\big)\wedge \,\overline{\!\theta}^{\hspace{0.2ex} j}_{(s-r)x}\,.
\label{theta}
\end{align}
In practice, it is often easier to determine $\vartheta$ from~\eqref{thetaequation} directly by inspection.

There are no other solutions to~\eqref{thetaequation} because the Lie derivative $\mathcal{L}_{\,\overline{\!D}_x}$ increases the maximum value of the index of the second tensor factor:
$$
\mathcal{L}_{\,\overline{\!D}_x}(\beta\otimes \,\overline{\!\theta}^{\hspace{0.2ex} j}_{sx}) = \mathcal{L}_{\,\overline{\!D}_x}(\beta)\otimes \,\overline{\!\theta}^{\hspace{0.2ex} j}_{sx} + \beta\otimes \,\overline{\!\theta}^{\hspace{0.2ex} j}_{(s+1)x}\qquad \text{for}\quad \beta\in \mathcal{C}\Lambda^1(\mathcal{E})\,.
$$
This means that the Lie derivative $\mathcal{L}_{\,\overline{\!D}_x}$ of a nonzero element of $\mathcal{C}^2\Lambda^2(\mathcal{E})\subset \mathcal{C}\Lambda^1(\mathcal{E})\otimes \mathcal{C}\Lambda^1(\mathcal{E})$ cannot be zero. Then the kernel of $dx\wedge\mathcal{L}_{\,\overline{\!D}_x}$ in~\eqref{thetaequation} is trivial.

Since~\eqref{thetaequation} has a unique solution, the remaining relation
\begin{align*}
\gamma_{ij2}^{ks}\ \overline{\!\theta}^{\hspace{0.2ex} i}_{kx}\wedge \,\overline{\!\theta}^{\hspace{0.2ex} j}_{sx}\wedge dt = dt\wedge\mathcal{L}_{\,\overline{\!D}_t}\vartheta\,,\qquad \,\overline{\!D}_t = D_t|_{\mathcal{E}}
\end{align*}
is satisfied automatically, and $\vartheta$ from~\eqref{theta} is a solution to the equation $\mathcal{L}_X\omega = d_0\vartheta$.

\vspace{1ex}

\remarka{This algorithm does not involve the $dt$-component of $\omega$ and can be directly generalized to the case of elements of an arbitrary group $E_1^{\hspace{0.1ex}p, \hspace{0.2ex} 1}$, $p\geqslant 1$
of a (1+1)-dimensional evolution system.}

\vspace{1ex}

\remarka{If $\mathcal{E}_X$ is a finite-dimensional smooth manifold and the restrictions of $\vartheta|_{\mathcal{E}_X}$ to fibers of $\pi_{\mathcal{E}_X}$ are invertible, then the differential $2$-form $\vartheta|_{\mathcal{E}_X}$ gives rise to a Poisson bivector. It seems that, under rather general assumptions regarding conservation laws of (1+1)-dimensional evolution systems with invertible presymplectic operators, such bivectors can be obtained through the method presented in Refs.~\cite{mokhov1984hamiltonian, mokhov1988hamiltonian}, utilizing the appropriate conservation laws (see also Ref.~\cite{ugaglia2002hamiltonian}). The exact relationship between these two approaches requires further investigation.
}

\section{Examples \label{Examplessec}}

In the examples below, we use more suitable index notation, as follows. Let us recall that $u^i$, $u^i_{x^j}$, $u^i_{x^jx^k} = u^i_{x^j+x^k}$, $\ldots$ denote coordinates on $J^{\infty}(\pi)$ (or on $\mathcal{E}$); in particular, they are not considered to be functions of independent variables.

\subsection{Reduction of conservation laws}\label{sec:red:cl}

Let us consider two examples. One of these examples deals with a point symmetry, whereas the other one concerns a higher symmetry.

\vspace{1ex}

\examplea{\label{Examplenonldif} Let us consider the (1+2)-dimensional nonlinear diffusion equation given by
$$
u_t = \mathrm{div}(u\,\mathrm{grad}\hspace{0.1ex} u).
$$
Here $u^1 = u$, $x^2 = x$, $x^3 = y$, $f^1 = u(u_{xx} + u_{yy}) + u_x^2 + u_y^2$, $F = F^1 = u_t - f^1$. This equation can be written in the total divergence form
$$
D_t(u) + D_x(-uu_x) + D_y(-uu_y) = 0,
$$
and hence admits the conservation law represented by
\begin{align}
\omega = u\, dx\wedge dy + uu_x dt\wedge dy - uu_y dt\wedge dx\,.
\label{diffuCL}
\end{align}
In \eqref{CLcharform}, we can put $A_1 = 0$, because
$$
d_h\hspace{0.2ex} \omega = F dt\wedge dx\wedge dy,
$$
and the integration by parts form \eqref{CLcharform} is not required. According to \eqref{cosymcompon},
the corresponding cosymmetry $\psi$ has the component $\psi_1 = 1$.

The conservation law is invariant under the action of the scaling point symmetry
$$
Y = -x\partial_x - y\partial_y - 4t\partial_t + 2u\partial_u,
$$
because $E_{\varphi}(\psi) + l_{\varphi}^{\,*}(\psi) = 0$, where $\varphi\in\varkappa(\mathcal{E})$ given by
$$
\varphi = 2u + xu_x + yu_y + 4t\big(u(u_{xx} + u_{yy}) + u_x^2 + u_y^2\big)
$$
is the characteristic of this symmetry. In components, integration by parts \eqref{intparts} yields
\begin{align*}
\psi_1\cdot l_{\varphi}(F) &= (2 + 4tu_{xx} + 4tu_{yy})F + (x + 8tu_x)D_x(F) + (y + 8tu_y)D_y(F) + 4tuD_x^2(F) + 4tuD_y^2(F)\\
&= D_x(xF + 4D_x(tuF)) + D_y(yF + 4D_y(tuF)) + F\cdot (\ldots)
\end{align*}
The last term corresponds to $\langle l_{\varphi}^{\,*}(\psi), F \rangle$, and does not contribute to $A_2 F$.
We obtain
$$
A_2 F = -(xF + 4D_x(tuF))dt\wedge dy + (yF + 4D_y(tuF))dt\wedge dx\,.
$$
The Lie derivative reads
$$
\mathcal{L}_{E_{\varphi}}\omega = \varphi\, dx\wedge dy + (\varphi u_x + uD_x(\varphi)) dt\wedge dy - (\varphi u_y + uD_y(\varphi)) dt\wedge dx\,.
$$
Applying the horizontal homotopy formula to the difference $\mathcal{L}_{E_{\varphi}}\omega - A_2F$ (i.e., the left-hand side of \eqref{algoformula}), one finds
\begin{align}
\widehat{\vartheta} = u(xu_y - yu_x)dt - u(y + 4tu_y)dx + u(x + 4tu_x)dy\,.
\label{clreduc}
\end{align}
The restriction of $\widehat{\vartheta}$ to the corresponding system $\mathcal{E}_X$ (given by the equations $F = 0$, $\varphi = 0$, and their differential consequences) represents the reduction of the conservation law represented by~\eqref{diffuCL}. This reduction can be written in the total curl form on the jets: the relation
\begin{align*}
\big(D_x(P_2) - D_y(P_1), D_y(P_0) - D_t(P_2), D_t(P_1) - D_x(P_0)\big) = (0, 0, 0)
\end{align*}
holds on $X$-invariant solutions for $P_0 = u(xu_y - yu_x)$, $P_1 = - u(y + 4tu_y)$, $P_2 = u(x + 4tu_x)$.

Using an $X$-invariant solution, one can integrate~\eqref{clreduc} over a closed (compact and without boundary) oriented 1-dimensional submanifold of the solution’s domain. The resulting integral is independent of the specific representative chosen for the reduction and remains conserved, depending solely on the homology class of the submanifold.
}

\vspace{1ex}

\remarka{The form $\widehat{\vartheta}$ turns out scaling-invariant. One can introduce the variables $w = u\sqrt{t}$, $\mu = x/\sqrt[4]{t}$, $\nu = y/\sqrt[4]{t}$, $\tau = \ln t$ assuming $t > 0$. Then the system $\mathcal{E}_X$ takes the form
$$
w_{\tau} = 0\,,\qquad -\dfrac{1}{4}(2w + \mu w_{\mu} + \nu w_{\nu}) = w(w_{\mu\mu} + w_{\nu\nu}) + w_{\mu}^2 + w_{\nu}^2\,,
$$
while
$$
\widehat{\vartheta} = - w(\nu + 4w_{\nu})d\mu + w(\mu + 4w_{\mu})d\nu\,.
$$
In these local coordinates, solutions to $\mathcal{E}_X$ are functions of the two new variables $\mu$ and $\nu$. Since $Y \lrcorner\, \widehat{\vartheta} = 0$, the scale-invariant form $\widehat{\vartheta}$ produces a conservation law of the quotient system (which doesn't involve $\tau$). It can be written in the total divergence form
\begin{align*}
D_{\mu}(w(\mu + 4w_{\mu})) + D_{\nu}(w(\nu + 4w_{\nu})) = 0\,.
\end{align*}
Let us note that~\eqref{clreduc} is written in global physical variables and plays the same role. In particular, one can introduce a nonlocal variable $b$ (potential) by the formula $d_0 b = \widehat{\vartheta}|_{\mathcal{E}_X}$. The relations
\begin{align*}
b_t = u(xu_y - yu_x)\,,\qquad b_x = - u(y + 4tu_y)\,,\qquad b_y = u(x + 4tu_x)\,.
\end{align*}
together with the equations $F = 0$ and $\varphi = 0$ determine a one-dimensional differential covering~\cite{VinKr} (also called a \emph{potential system}, see, e.g., Ref.~\cite{bluman2010applications}) of the system $\mathcal{E}_X$.
}

\vspace{1ex}

\remarka{
In the general case, one can use elements of $E_1^{\hspace{0.1ex}0, \hspace{0.1ex} 1}$ to introduce nonlocal variables (potentials) resulting in differential coverings with finite-dimensional fibers.
}

\vspace{1ex}

\examplea{\label{ExampleCBS} Consider the infinite prolongation $\mathcal{E}$ of the Calogero--Bogoyavlenskii--Schiff breaking soliton equation
\begin{align}
u_{tx} = 2u_y u_{xx} + 4u_x u_{xy} - u_{xxxy}\,.
\label{CBS}
\end{align}
The equation is remarkable for many reasons \cite{Bogoyavlenskii_1990, Schiff:1990sa, KRASILSHCHIK2023104927}. In particular, it is an S-integrable Lagrangian equation that admits recursion operators generating an infinite hierarchy of local higher symmetries~\cite{KRASILSHCHIK2023104927}. As coordinates on $\mathcal{E}$, we take all the variables except $u_{tx}$ and its total derivatives.

Let us examine the conservation laws represented by $\omega_1, \omega_2\in E_0^{\hspace{0.1ex}0, \hspace{0.1ex} 2}(\mathcal{E})$,
\begin{align*}
&\omega_1 = (u_{xxx} - u_x^2)dt\wedge dx + 2u_x u_y dt\wedge dy + u_x dx\wedge dy\,,\\
&\omega_2 = \Big(u_x u_{xxx} + \dfrac{u_{xx}^2}{2} - u_x^3\Big)dt\wedge dx + (u_x^2u_y + u_{xx}u_{xy})dt\wedge dy + \dfrac{u_x^2}{2} dx\wedge dy\,.
\end{align*}
On the jets, they can be written in the total divergence form
\begin{align*}
&D_t(u_x) + D_x(-2u_x u_y) + D_y(u_{xxx} - u_x^2) = 0\,,\\
&D_t\Big(\dfrac{u_x^2}{2}\Big) + D_x(-(u_x^2u_y + u_{xx}u_{xy})) + D_y\Big(u_x u_{xxx} + \dfrac{u_{xx}^2}{2} - u_x^3\Big) = 0\,,
\end{align*}
respectively.
These conservation laws are invariant with respect to a higher symmetry $X = E_{\varphi}|_{\mathcal{E}}$ with
\begin{align*}
\varphi = u_{xxx} - 3u_x^2\,.
\end{align*}
This follows from the observation that the cosymmetries of the conservation laws represented by $\mathcal{L}_X \omega_i$ are zero ($i = 1, 2$).

Directly solving the equations $\mathcal{L}_X \omega_i = d_0 \vartheta_i$ on $\mathcal{E}$, we find the two-component horizontal $1$-forms
\begin{align*}
&\vartheta_1 = \varphi\, dy - (u_{5x} - 8u_x u_{xxx} - 5u_{xx}^2 + 4u_x^3)dt\,,\\
&\vartheta_2 = \Big(u_x u_{xxx} - \dfrac{u_{xx}^2}{2} - 2u_x^3\Big)dy - \Big(u_x u_{5x} + \dfrac{u_{xxx}^2}{2} - 9u_x^2 u_{xxx} - 6u_x u_{xx}^2 + \dfrac{9}{2}u_x^4\Big)dt\,.
\end{align*}
All other solutions differ from these by adding terms of the form $d_0 f$, where $f$ is a function on $\mathcal{E}$.
Moreover, the restrictions of these $\vartheta_i$ to the system $\mathcal{E}_X$ (given by the equations $\varphi = 0$, \eqref{CBS}, and their differential consequences) are the one-component horizontal $1$-forms
\begin{align*}
\vartheta_1|_{\mathcal{E}_X} = 2g\hspace{0.15ex} dt\,,\qquad
\vartheta_2|_{\mathcal{E}_X} = g\hspace{0.15ex} dy\,, \qquad g = u_x^3 - \dfrac{u_{xx}^2}{2}\,,
\end{align*}
representing elements of $E_1^{\hspace{0.1ex}0, \hspace{0.2ex} 1}(\mathcal{E}_X)$. Consequently, the function $g$
is a constant of $X$-invariant motion, $g\in E_1^{\hspace{0.1ex}0, \hspace{0.2ex} 0}(\mathcal{E}_X)$. In other words, for every $X$-invariant solution (whether global or local, with a connected domain $\subset \mathbb{R}^3$), there exists a constant $C \in \mathbb{R}$ such that the relation $g = C$ holds for the solution.
}

\vspace{1ex}

\remarka{Equation~\eqref{CBS} is $\ell$-normal (and has trivial de Rham cohomology groups $H_{dR}^i$ for $i > 0$). Therefore, its group $E_1^{\hspace{0.1ex}0, \hspace{0.2ex} 1}$ is trivial, and $g$ cannot be obtained directly through the reduction mechanism.
}

\subsection{Reduction of presymplectic structures}\label{sec:red:pre}

Let us consider three examples, two of which rely on the algorithm from Section~\ref{Redofpresstr}. One of these examples pertains to a point symmetry, while the others involve higher symmetries.

\vspace{1ex}

\examplea{\label{Examplewave} Let us consider an example of a Lagrangian system and a point Noether symmetry. One can rewrite the equation
$$
u_{tt} = (1 + u_x^2)u_{xx}
$$
in the evolution form
\begin{align}
&u_t = v\,,\qquad v_t = (1 + u_x^2)u_{xx}\,.
\label{wave}
\end{align}
Here $u^1 = u$, $u^2 = v$, $f^1 = v$, $f^2 = (1 + u_x^2)u_{xx}$, $F^1 = u_t - f^1$, $F^2 = v_t - f^2$.
The variational derivative $\mathrm{E}(L)$ of the Lagrangian
\begin{align*}
L = \lambda\, dt\wedge dx\,,\qquad \lambda = \dfrac{v^2}{2} + \dfrac{u_x^2}{2} +\dfrac{u_x^4}{12} - u_t v\,,
\end{align*}
vanishes on the infinite prolongation $\mathcal{E}$ of~\eqref{wave}:
\begin{align*}
\dfrac{\delta \lambda}{\delta u} = F^2\,,\qquad \dfrac{\delta \lambda}{\delta v} = -F^1\,.
\end{align*}
Then we can take the operator determined by the matrix
\begin{align*}
\begin{pmatrix}
0 & 1\\
-1 & 0
\end{pmatrix}
\end{align*}
as $B$ in~\eqref{variop}. The presymplectic operator $\Delta = B^*|_{\mathcal{E}}$ maps the characteristic $\widetilde{\varphi} = (\widetilde{\varphi}^{\hspace{0.2ex} 1}, \widetilde{\varphi}^{\hspace{0.2ex} 2}) \in \varkappa(\mathcal{E})$ of a symmetry of $\mathcal{E}$ to the cosymmetry that has the components $ (-\widetilde{\varphi}^{\hspace{0.2ex} 2}, \widetilde{\varphi}^{\hspace{0.2ex} 1})$.

The Lie derivative $\mathcal{L}_{E_{\chi}}(L)$ from~\eqref{Noethiden} reads
\begin{align*}
\mathcal{L}_{E_{\chi}}(L) = \Big(v\chi^2 + u_xD_x(\chi^1) + \dfrac{u_x^3}{3}D_x(\chi^1) - u_t\chi^2 - vD_t(\chi^1)\Big) dt\wedge dx
\end{align*}
for $\chi\in \varkappa(\pi)$. Integrating by parts, we get the Noether identity~\eqref{Noethiden}
\begin{align*}
\mathcal{L}_{E_{\chi}}(L) = \langle \mathrm{E}(L), \chi \rangle + D_t(-v\chi^1)dt\wedge dx + D_x(\ldots)dt\wedge dx\,.
\end{align*}
Then one can choose the following presymplectic potential current
\begin{align*}
\omega_L = -v\, \theta^1\wedge dx + \ldots \wedge dt\,,
\end{align*}
and the presymplectic structure is represented by the differential form $\omega\in E^{\hspace{0.1ex}2, \hspace{0.2ex} 1}_0(\mathcal{E})$,
\begin{align*}
\omega = d_v\hspace{0.15ex} \omega_L|_{\mathcal{E}} = -\,\overline{\!\theta}^2\wedge \,\overline{\!\theta}^1\wedge dx + \ldots \wedge dt\,,\qquad \,\overline{\!\theta}^i_{\alpha} = \theta^i_{\alpha}|_{\mathcal{E}}\,.
\end{align*}

The PDE system~\eqref{wave} admits the point symmetry $\partial_x + t\partial_u + \partial_v$. Its characteristic $\varphi$ has the components
$$
\varphi^1 = t - u_x\,,\qquad \varphi^2 = 1 - v_x\,.
$$
The presymplectic structure is $X$-invariant (where $X = E_{\varphi}|_{\mathcal{E}}$), since $l_{\psi} - l^{\, *}_{\psi} = 0$ for $\psi = \Delta(\varphi)$. This is equivalent to the fact that $X$ is a Noether symmetry (i.e., it corresponds to a conservation law), since $\mathcal{E}$ is an evolution system, and hence, $\ell$-normal. The Lie derivative $\mathcal{L}_X \hspace{0.2ex} \omega$ has the form
\begin{align*}
\mathcal{L}_X \hspace{0.2ex} \omega = \,\overline{\!\theta}^2_x \wedge \,\overline{\!\theta}^1 \wedge dx + \,\overline{\!\theta}^2 \wedge \,\overline{\!\theta}^1_x \wedge dx + \ldots\wedge dt = dx\wedge \mathcal{L}_{\,\overline{\!D}_x}(\,\overline{\!\theta}^2 \wedge \,\overline{\!\theta}^1) + \ldots\wedge dt\,.
\end{align*}
Therefore, $\mathcal{L}_X \hspace{0.2ex} \omega = d_0 \vartheta$ for $\vartheta = \,\overline{\!\theta}^2 \wedge \,\overline{\!\theta}^1$.

One can use the variables $t$, $x$, $u$, $v$ as coordinates on $\mathcal{E}_X$, which is the infinite prolongation of
$$
u_t = v\,,\qquad u_x = t\,,\qquad v_t = 0\,,\qquad v_x = 1\,.
$$
The resulting reduction of the presymplectic structure is $\vartheta|_{\mathcal{E}_X}$, where
\begin{align*}
\vartheta|_{\mathcal{E}_X} = \,\tilde{\!\theta}^2 \wedge \,\tilde{\!\theta}^1,\qquad \,\tilde{\!\theta}^1 = \,\overline{\!\theta}^1|_{\mathcal{E}_X} = du - vdt - t dx\,,\qquad \,\tilde{\!\theta}^2 = \,\overline{\!\theta}^2|_{\mathcal{E}_X} = dv - dx\,.
\end{align*}
}

\vspace{1ex}

\examplea{\label{ExamKB} Let us consider the invariant reduction of a presymplectic structure of the potential Kaup--Boussinesq system
\begin{align}
v_t = -\dfrac{v_x^2}{2} - \eta_x\,,\qquad \eta_t = -v_x\eta_x -\dfrac{1}{4}v_{xxx}\,.
\label{KBClebsch}
\end{align}
Here $u^1 = v$, $u^2 = \eta$, $F^1 = v_t + v_x^2/2 + \eta_x$, $F^2 = \eta_t + v_x\eta_x + v_{xxx}/4$. This system is a two-dimensional differential covering of the Kaup--Boussinesq equations
$$
u_t + uu_x + h_x = 0, \qquad h_t + (hu)_x + \dfrac{1}{4}u_{xxx} = 0.
$$
The covering is determined by the Clebsch potentials $(v, \eta)$ satisfying $u = v_x$, $h = \eta_x$.

The variational derivative $\mathrm{E}(L)$ of the Lagrangian
\begin{align*}
L = \lambda\, dt\wedge dx\,,\qquad \lambda = -\dfrac{1}{2}\Big(v_t\eta_x + v_x\eta_t + v_x^2\eta_x + \eta_x^2 + \dfrac{1}{4}v_x v_{xxx}\Big)
\end{align*}
vanishes on the infinite prolongation $\mathcal{E}$ of~\eqref{KBClebsch}:
\begin{align*}
\dfrac{\delta \lambda}{\delta v} = D_x(F^2)\,,\qquad \dfrac{\delta \lambda}{\delta \eta} = D_x(F^1)\,.
\end{align*}
Then we can take the operator determined by the matrix
$$
\begin{pmatrix}
0 & D_x\\
D_x & 0
\end{pmatrix}
$$
as $B$ in~\eqref{variop}.
The presymplectic operator $\Delta = B^*|_{\mathcal{E}}$ maps the characteristic $\widetilde{\varphi} = (\widetilde{\varphi}^{\hspace{0.2ex} 1}, \widetilde{\varphi}^{\hspace{0.2ex} 2}) \in \varkappa(\mathcal{E})$ of a symmetry of $\mathcal{E}$ to the cosymmetry that has the components $ (-\,\overline{\!D}_x(\widetilde{\varphi}^{\hspace{0.2ex} 2}), -\,\overline{\!D}_x(\widetilde{\varphi}^{\hspace{0.2ex} 1}))$.

Consider the symmetry $X = E_{\varphi}|_{\mathcal{E}}$, where the characteristic $\varphi$ is given by its components
\begin{align*}
\varphi^1 = \dfrac{1}{3}v_{xxx} + 2v_x\eta_x + \dfrac{1}{3}v_x^3\,,\qquad \varphi^2 = \dfrac{1}{3}\eta_{xxx} + \dfrac{1}{2}v_x v_{xxx} + \dfrac{1}{4}v_{xx}^2 + v_x^2\eta_x + \eta_x^2\,.
\end{align*}
For $\psi = \Delta(\varphi)$, the presymplectic operator $l_{\psi} - l^{\,*}_{\psi}$ is zero (i.e., $\Delta(\varphi)$ is the cosymmetry of a conservation law). Then the corresponding presymplectic structure is $X$-invariant.

The Lie derivative $\mathcal{L}_{E_{\chi}}(L)$ from the Noether identity~\eqref{Noethiden} reads
\begin{align*}
\mathcal{L}_{E_{\chi}}(L) = -\dfrac{1}{2}\Big(&D_t(\chi^1)\eta_x + v_t D_x(\chi^2) + D_x(\chi^1)\eta_t + v_x D_t(\chi^2) + 2v_x D_x(\chi^1)\eta_x + v_x^2 D_x(\chi^2) {}\\
&+ 2\eta_xD_x(\chi^2) + \dfrac{1}{4}D_x(\chi^1) v_{xxx} + \dfrac{1}{4}v_x D_{3x}(\chi^1)\Big) dt\wedge dx
\end{align*}
for $\chi\in \varkappa(\pi)$. Integrating by parts, we get the Noether identity
\begin{align*}
\mathcal{L}_{E_{\chi}}(L) = \langle \mathrm{E}(L), \chi \rangle + D_t\Big(\dfrac{-\eta_x\chi^1 - v_x\chi^2}{2}\Big)dt\wedge dx + D_x(\ldots)dt\wedge dx\,.
\end{align*}
Then one can choose the following presymplectic potential current in~\eqref{Noethiden}
\begin{align*}
\omega_L = -\dfrac{1}{2}\eta_x\hspace{0.15ex} \theta^1\wedge dx - \dfrac{1}{2}v_x\hspace{0.15ex} \theta^2\wedge dx + \ldots \wedge dt\,,
\end{align*}
and the presymplectic structure is represented, e.g., by the differential form $\omega\in E^{\hspace{0.1ex}2, \hspace{0.2ex} 1}_0(\mathcal{E})$,
\begin{align*}
\omega = d_v\hspace{0.15ex} \omega_L|_{\mathcal{E}} + d_0\Big(\dfrac{1}{2}\,\overline{\!\theta}^2\wedge \,\overline{\!\theta}^1\Big) = -\,\overline{\!\theta}^1_x\wedge \,\overline{\!\theta}^2\wedge dx + \ldots \wedge dt\,.
\end{align*}
Its Lie derivative reads
\begin{align*}
\mathcal{L}_{X}\hspace{0.2ex} \omega = & -\Big(\dfrac{1}{3} \,\overline{\!\theta}^{\hspace{0.1ex} 1}_{xxxx} + 2v_{xx}\,\overline{\!\theta}^{\hspace{0.1ex} 2}_{x} + 2\eta_x \,\overline{\!\theta}^{\hspace{0.1ex} 1}_{xx} + 2v_{x}\,\overline{\!\theta}^{\hspace{0.1ex} 2}_{xx} + 2\eta_{xx} \,\overline{\!\theta}^{\hspace{0.1ex} 1}_{x} + v_x^2 \,\overline{\!\theta}^{\hspace{0.1ex} 1}_{xx} + 2v_x v_{xx}\,\overline{\!\theta}^{\hspace{0.1ex} 1}_{x}\Big)\wedge \,\overline{\!\theta}^{\hspace{0.1ex} 2}\wedge dx{}\\
& - \,\overline{\!\theta}^{\hspace{0.1ex} 1}_x \wedge \Big(\dfrac{1}{3} \,\overline{\!\theta}^{\hspace{0.1ex} 2}_{xxx} + \dfrac{1}{2}v_x \,\overline{\!\theta}^{\hspace{0.1ex} 1}_{xxx} + \dfrac{1}{2}v_{xx} \,\overline{\!\theta}^{\hspace{0.1ex} 1}_{xx} + (v_x^2 + 2\eta_x) \,\overline{\!\theta}^{\hspace{0.1ex} 2}_{x}\Big)\wedge dx + \ldots\wedge dt\,.
\end{align*}
Applying integration by parts, we find that $\mathcal{L}_X \omega = d_0 \vartheta$ for
\begin{align*}
\vartheta = -\dfrac{1}{3} \,\overline{\!\theta}^{\hspace{0.1ex} 1}_{xxx}\wedge \,\overline{\!\theta}^{\hspace{0.1ex} 2} + \dfrac{1}{3} \,\overline{\!\theta}^{\hspace{0.1ex} 1}_{xx}\wedge \,\overline{\!\theta}^{\hspace{0.1ex} 2}_{x} - \dfrac{1}{3} \,\overline{\!\theta}^{\hspace{0.1ex} 1}_{x}\wedge \,\overline{\!\theta}^{\hspace{0.1ex} 2}_{xx} - \,\overline{\!\theta}^{\hspace{0.1ex} 1}_{x}\wedge \dfrac{1}{2} v_x\,\overline{\!\theta}^{\hspace{0.1ex} 1}_{xx} - (v_x^2 + 2\eta_x)\,\overline{\!\theta}^{\hspace{0.1ex} 1}_{x}\wedge \,\overline{\!\theta}^{\hspace{0.1ex} 2} - 2v_x \,\overline{\!\theta}^{\hspace{0.1ex} 2}_{x}\wedge \,\overline{\!\theta}^{\hspace{0.1ex} 2}.
\end{align*}
The invariant system $\mathcal{E}_X$ is given by the infinite prolongation of equations~\eqref{KBClebsch} and
$$
v_{xxx} = -6v_x\eta_x - v_x^3\,,\qquad \eta_{xxx} = 6v_x^2\eta_x + \dfrac{3}{2}v_x^4 - \dfrac{3}{4}v_{xx}^2 - 3\eta_x^2\,.
$$
Finally, reduction of the presymplectic structure yields
\begin{align*}
\vartheta|_{\mathcal{E}_X} = \dfrac{1}{3} \,\tilde{\!\theta}^{\hspace{0.1ex} 1}_{xx}\wedge \,\tilde{\!\theta}^{\hspace{0.1ex} 2}_{x} - \dfrac{1}{3} \,\tilde{\!\theta}^{\hspace{0.1ex} 1}_{x}\wedge \,\tilde{\!\theta}^{\hspace{0.1ex} 2}_{xx} - \dfrac{1}{2}v_x \,\tilde{\!\theta}^{\hspace{0.1ex} 1}_{x}\wedge \,\tilde{\!\theta}^{\hspace{0.1ex} 1}_{xx}\,, \qquad \,\tilde{\!\theta}^{\hspace{0.1ex} i}_{kx} = \,\overline{\!\theta}^{\hspace{0.1ex} i}_{kx}|_{\mathcal{E}_X}\,.
\end{align*}

Let us demonstrate how the Noether theorem for invariant solutions works. Substituting the symmetry
$Y = \partial_x$ of $\mathcal{E}_X$
to $\vartheta|_{\mathcal{E}_X}$, we obtain
\begin{align*}
Y \lrcorner\, (\vartheta|_{\mathcal{E}_X}) &= \Big(2v_x\eta_x + \dfrac{1}{3}v_x^3\Big)\,\tilde{\!\theta}^{\hspace{0.1ex} 2}_{x} + \dfrac{1}{3}\eta_{xx}\,\tilde{\!\theta}^{\hspace{0.1ex} 1}_{xx} + \dfrac{1}{3}v_{xx}\,\tilde{\!\theta}^{\hspace{0.1ex} 2}_{xx} + \Big(\dfrac{1}{4}v_{xx}^2 + v_x^2\eta_x + \eta_x^2\Big)\,\tilde{\!\theta}^{\hspace{0.1ex} 1}_{x} + \dfrac{1}{2}v_x v_{xx} \,\tilde{\!\theta}^{\hspace{0.1ex} 1}_{xx} {}\\
&= d\Big(\dfrac{1}{3}v_{xx}\eta_{xx} + v_x\eta_x^2 + \dfrac{1}{3}v_x^3\eta_x + \dfrac{1}{4}v_x v_{xx}^2\Big)\,.
\end{align*}
Then the symmetry $Y$ corresponds to the constant of $X$-invariant motion
$$
\dfrac{1}{3}v_{xx}\eta_{xx} + v_x\eta_x^2 + \dfrac{1}{3}v_x^3\eta_x + \dfrac{1}{4}v_x v_{xx}^2\,.
$$
}

\remarka{(On presymplectic reduction). Taking the quotient by the group action $v \mapsto v + \epsilon_1$, $\eta \mapsto \eta + \epsilon_2$ on $\mathcal{E}_X$ (with group parameters $\epsilon_1, \epsilon_2$) and abusing notation, we get the differential covering $(t, x, v, \eta, v_x, \eta_x, v_{xx}, \eta_{xx}) \mapsto (t, x, v_x, \eta_x, v_{xx}, \eta_{xx})$ from $\mathcal{E}_X$ to the quotient system. The Cartan distribution of the quotient system is spanned by the (well-defined) projections of the total derivatives from $\mathcal{E}_X$. Then $\vartheta|_{\mathcal{E}_X}$ is the lift of the closed differential $2$-form that has the same expression in the coordinates on the quotient system. This $2$-form is non-degenerate on fibers of the quotient system bundle $\widetilde{\pi}_{\mathcal{E}_X}\colon (t, x, v_x, \eta_x, v_{xx}, \eta_{xx}) \mapsto (t, x)$ and consequently gives rise to a Poisson bracket\footnote{The situation with the presymplectic reduction (quotient by the kernel) is more complicated for presymplectic structures of PDEs, as they are genuine cohomology classes. For presymplectic reduction of finite-dimensional smooth manifolds and Hamiltonian systems see, e.g., Ref.~\cite{doi:10.1142/S0129055X99000386}.}. Informally speaking, the symmetries $\partial_x$ and $\partial_t$ of the quotient system result in its Liouville integrability, as in Remark~\ref{ThelastRem}.

}

\vspace{1ex}

\examplea{Let us take as $\mathcal{E}$ the cotangent covering of the $r^{\rm th}$ dispersionless Dym equation (see, e.g., Refs.~\cite{blaszak2002classical, pavlov2003integrable, morozov2009contact, ovsienko2010bi}) \label{ExamplerDym} given by the infinite prolongation of
\begin{align}
v_{ty} = u_x v_{xy} - u_y v_{xx} + 2(u_{xx} v_y - u_{xy} v_x)\,, \qquad u_{ty} = u_x u_{xy} - u_y u_{xx}\,.
\label{rDymcot}
\end{align}
Here $u^1 = u$, $u^2 = v$, The system~\eqref{rDymcot} coincides with the Euler--Lagrange equation for
\begin{align*}
L = v(u_{ty} - u_x u_{xy} + u_y u_{xx})\, dt\wedge dx\wedge dy\,.
\end{align*}
It admits the Noether symmetry $X = E_{\varphi}|_{\mathcal{E}}$, where $\varphi$ is given by its components~\cite{baran2014higher}
\begin{align*}
\varphi^1 = 0\,,\qquad \varphi^2 = u_{txx} - u_x u_{xxx} - \dfrac{1}{2}u_{xx}^2\,.
\end{align*}
The presymplectic structure induced by $L$ is $X$-invariant. Let us denote $\theta = du - u_t dt - u_x dx - u_y dy$, $\theta_t = du_t - u_{tt} dt - u_{tx} dx - u_{ty} dy$, $\ldots$ One can choose the presymplectic potential current in~\eqref{Noethiden}
\begin{align*}
\omega_L =\ &(v \theta_t - u_x v \theta_x + u_{xx} v\theta)\wedge dt\wedge dx + \big((u_y v_x - u_{x}v_y + u_{xy}v)\theta - u_y v \theta_x\big)\wedge dt\wedge dy\\
&- v_y\theta \wedge dx\wedge dy\,.
\end{align*}
Then for $\omega = d_v\hspace{0.15ex} \omega_L|_{\mathcal{E}}$, one finds that $\mathcal{L}_X \omega = d_0 \vartheta$,
\begin{align*}
\vartheta &= \Big((u_x \,\overline{\!\theta}_{tx} - u_x^2 \,\overline{\!\theta}_{xx})\wedge \,\overline{\!\theta}_x + (-\,\overline{\!\theta}_{tx} + u_{xx}\,\overline{\!\theta}_x + u_x \,\overline{\!\theta}_{xx})\wedge \,\overline{\!\theta}_t + \dfrac{1}{2}(2u_x u_{xx} \,\overline{\!\theta}_{xx} + u_{xx}^2 \,\overline{\!\theta}_x)\wedge \,\overline{\!\theta}\Big) \wedge dt\\
&+ u_{xx}\,\overline{\!\theta}_{xx}\wedge \,\overline{\!\theta}\wedge dx + \big((-u_y \,\overline{\!\theta}_{xx} - u_{xx} \,\overline{\!\theta}_{y})\wedge \,\overline{\!\theta}_x + (u_{xx} \,\overline{\!\theta}_{xy} + u_{xy} \,\overline{\!\theta}_{xx} + u_y \,\overline{\!\theta}_{xxx} + u_{xxx} \,\overline{\!\theta}_{y})\wedge \,\overline{\!\theta}\big)\wedge dy\,.
\end{align*}
Here $\,\overline{\!\theta} = \theta|_{\mathcal{E}}$, $\,\overline{\!\theta}_t = \theta_t|_{\mathcal{E}}$, $\ldots$ It is important to note that $\vartheta$ can be interpreted as a differential form on the infinite prolongation of the $r^{\rm th}$ dispersionless Dym equation itself. Its restriction to the (infinitely prolonged) system\footnote{It is regular, for example, if one assumes $u_y \neq 0$. More formally, one can introduce an auxiliary dependent variable $w\in\mathbb{R}\setminus \{0\}$ and add the equation $w = u_y$.}
\begin{align*}
u_{ty} = u_x u_{xy} - u_y u_{xx}\,,\qquad u_{txx} - u_x u_{xxx} - \dfrac{1}{2}u_{xx}^2 = 0
\end{align*}
defines an element of its group $E_1^{2, 1}$ and yields the corresponding Noether theorem as in Section~\ref{NoetherTheo}. A compatibility complex for the linearization of this system allows one to describe the Noether theorem in a more convenient form -- in terms of total differential operators.

\vspace{1ex}

\remarka{The system~\eqref{rDymcot} admits the following point symmetries
\begin{align*}
Y_1 = t\partial_t + x\partial_x + u\partial_u - 2v\partial_v + \ldots\,,\qquad Y_2 = x\partial_x + 2u\partial_u - 3v\partial_v + \ldots
\end{align*}
Both of them are Noether symmetries; $Y_1$ commutes with $X$, whereas $[X, Y_2] = -3X$. Since they are restrictions of symmetries (of the corresponding infinite jet space) that generate flows preserving $\mathcal{E}_X$, the symmetries $Y_1$ and $Y_2$ restrict to $\mathcal{E}_X$. One can see that the element of $E_1^{2, 1}(\mathcal{E}_X)$ defined by $\vartheta|_{\mathcal{E}_X}$ is $Y_1|_{\mathcal{E}_X}$-invariant but not $Y_2|_{\mathcal{E}_X}$-invariant (this can be considered an illustration of Proposition~\ref{Proposition}). From a global perspective, its reduction\footnote{We do not address the question of well-definedness of the reduction under $Y_1|_{\mathcal{E}_X}$.} under $Y_1|_{\mathcal{E}_X}$ can be based on the mechanism in the form given in Appendix~\ref{App:A}. Since the form $\vartheta|_{\mathcal{E}_X}$ is invariant under the flow of the vector field $Y_1|_{\mathcal{E}_X}$, the reduction is represented by the restriction of $-Y_1|_{\mathcal{E}_X} \lrcorner\, \vartheta|_{\mathcal{E}_X}$ to $\mathcal{E}_{X, Y_1}$, which coincides with the restriction of $-(t\,\overline{\!D}_t + x\,\overline{\!D}_x) \lrcorner\, \vartheta$ to $\mathcal{E}_{X, Y_1}$. Here, $Y_1$ restricts to the system $\mathcal{E}_{X, Y_1}$ given by the infinite prolongation of}
\begin{align*}
&v_{ty} = u_x v_{xy} - u_y v_{xx} + 2(u_{xx} v_y - u_{xy} v_x)\,, \qquad u_{ty} = u_x u_{xy} - u_y u_{xx}\,,\\
&u_{txx} - u_x u_{xxx} - \dfrac{1}{2}u_{xx}^2 = 0\,,\qquad u - tu_t - xu_x = 0\,,\qquad -2v - tv_t - xv_x = 0\,.
\end{align*}
}

\subsection{Reduction of variational principles}\label{sec:red:var}

Let us examine two examples of the reduction of the stationary action principle. One of these examples deals with a point symmetry, whereas the other one concerns a higher symmetry.

From a computational point of view, it is more convenient to describe reduction of presymplectic structures.
The generalization of Theorem~\ref{Theor1} (see Appendix~\ref{App:B}) shows that if a differential form $\varrho$ represents the reduction of an $X$-invariant internal Lagrangian, then $-d\varrho$ represents the reduction of the corresponding presymplectic structure provided that both these reductions are well-defined. This is the case, e.g., when $\mathcal{E}$ is $\ell$-normal.

\vspace{1ex}

\examplea{\label{Examplewavestatact} Let us demonstrate reduction of the variational principle for~\eqref{wave} from Example~\ref{Examplewave}.
The reduction of the internal Lagrangian from Example~\ref{Examplewave} is represented by any potential $\varrho$ for $-\vartheta|_{\mathcal{E}_X}$. For instance, $-\vartheta|_{\mathcal{E}_X} = d\varrho$, where
$$
\varrho = (u - tv)\,\tilde{\!\theta}^2 = (u - tv)(dv - dx).
$$
In coordinates, any section (not necessarily a solution) of $\pi_{\mathcal{E}_X}$ has the form
\begin{align*}
u = a(t, x)\,,\qquad v = b(t, x)\,.
\end{align*}
Choose a compact submanifold $N^1\subset \mathbb{R}^2$. In Definition~\ref{Def1}, it suffices to consider paths of the form
$$
\gamma(\tau)\colon \qquad u = a + \tau\delta a\,,\qquad v = b + \tau\delta b\,,
$$
where $a, b, \delta a, \delta b\in C^{\infty}(\mathbb{R}^2)$ are arbitrary such that $\delta a$ and $\delta b$ vanish on $\partial N$.
Then
$$
\gamma(\tau)^*(\varrho) = (a + \tau\delta a - t(b + \tau\delta b))(d(b + \tau\delta b) - dx)
$$
and~\eqref{varpr} takes the form
\begin{align*}
0 &= \dfrac{d}{d\tau}\Big|_{\tau = 0}\int_N \gamma(\tau)^*(\varrho) = \int_N (\delta a - t\delta b)(db - dx) + (a - tb)d(\delta b) \\
&= \int_N (\delta a - t\delta b)(db - dx) - \delta b\, d(a - tb)
= \int_N \delta a\, \sigma^*\tilde{\theta}^2 - \delta b\, \sigma^*\tilde{\theta}^1\,.
\end{align*}
for $\sigma = \gamma(0)$.
Because of the arbitrariness of $N$, $\delta a$, and $\delta b$, stationary points of the reduction of the internal Lagrangian are described by the equations $\sigma^*\tilde{\theta}^1 = \sigma^*\tilde{\theta}^2 = 0$ or, in other words, by
$$
d b = dx\,,\qquad d a = b dt + t dx\,.
$$
One can see that a section $\sigma$ is a stationary point if and only if it is a solution to $\mathcal{E}_X$.
}

\vspace{1ex}

\remarka{
The form $\varrho$ is invariant under the symmetry $\partial_x + t\partial_u + \partial_v$, and their interior product is trivial. One can introduce the variables $h = u - xt$, $k = v - x$. Then $\varrho = (h - tk) dk$ leads to an internal Lagrangian of the quotient system given by the equations
$$
h_t = k\,,\qquad k_t = 0
$$
for functions of the single variable $t$. Denote by $\widetilde{\pi}_{\mathcal{E}_X}\colon (t, h, k)\mapsto t$ the bundle of the quotient system. One can vary the internal Lagrangian of the quotient system within the class of all sections of $\widetilde{\pi}_{\mathcal{E}_X}$. This gives rise to the Lagrangian
$$
\widetilde{L} = (h - tk) k_t\, dt \qquad \text{on}\quad J^{1}(\widetilde{\pi}_{\mathcal{E}_X})\,.
$$
The corresponding Euler--Lagrange equations reproduce the quotient system.
}

\vspace{1ex}

\examplea{\label{ExampleNLS} Consider the infinite prolongation $\mathcal{E}$ of the nonlinear Schr{\"o}dinger equation (NLS)
\begin{align*}
u_t = -\dfrac{v_{xx}}{2} + (u^2 + v^2)v\,,\qquad v_t = \dfrac{u_{xx}}{2} - (u^2 + v^2)u\,.
\end{align*}
Using Noether's identity~\eqref{Noethiden}, one finds that the Lagrangian of the NLS
\begin{align}
L = -\dfrac{1}{2}\Big(u v_t - u_t v + \dfrac{u_x^2 + v_x^2}{2} + \dfrac{(u^2 + v^2)^2}{2}\Big) dt\wedge dx
\label{varprNLS}
\end{align}
gives rise to the internal Lagrangian represented by
\begin{align*}
l = -\dfrac{1}{4}\big(uu_{xx} + vv_{xx} + u_x^2 + v_x^2 - (u^2 + v^2)^2\big) dt\wedge dx - \dfrac{1}{2}(u\,\overline{\!\theta}^2 - v\,\overline{\!\theta}^1)\wedge dx + \dfrac{1}{2} (u_x\,\overline{\!\theta}^1 + v_x\,\overline{\!\theta}^2)\wedge dt\,,
\end{align*}
where $\,\overline{\!\theta}^1 = du - u_x dx - (-v_{xx}/2 + (u^2 + v^2)v)dt$ and $\,\overline{\!\theta}^2 = dv - v_x dx - (u_{xx}/2 - (u^2 + v^2)u)dt$. The corresponding presymplectic structure is represented by
\begin{align*}
\omega = dl = - \ \overline{\!\theta}^1\wedge \,\overline{\!\theta}^2\wedge dx +
\ldots \wedge dt\,.
\end{align*}
The NLS possesses the Noether symmetry $X = E_{\varphi}|_{\mathcal{E}}$ given by
\begin{align*}
\varphi^1 = u_{xxx} - 6(u^2 + v^2)u_x\,,\qquad \varphi^2 = v_{xxx} - 6(u^2 + v^2)v_x\,.
\end{align*}
The system $\mathcal{E}_X$ for $X$-invariant solutions is given by the NLS, equations $\varphi^1 = 0$, $\varphi^2 = 0$, and their differential consequences. We can regard $t$, $x$, $u$, $v$, $u_x$, $v_x$, $u_{xx}$, $v_{xx}$ as coordinates on it.

One has $\mathcal{L}_X \omega = d_0 \vartheta$ for
\[
\vartheta = -\ \overline{\!\theta}^1_{xx} \wedge \,\overline{\!\theta}^2 + \,\overline{\!\theta}^1_{x} \wedge \,\overline{\!\theta}^2_x - \,\overline{\!\theta}^1 \wedge \,\overline{\!\theta}^2_{xx} + 6(u^2 + v^2)\ \overline{\!\theta}^1\wedge \,\overline{\!\theta}^2,
\]
where 
\[
\overline{\!\theta}^i_{x} = \mathcal{L}_{\,\overline{\! D}_x} \,\overline{\!\theta}^i,\qquad \overline{\!\theta}^i_{xx} = \mathcal{L}_{\,\overline{\! D}_x} \,\overline{\!\theta}^i_x\,, \qquad \overline{\! D}_x = D_x|_{\mathcal{E}}.
\]
Then the reduction of the presymplectic structure is $\vartheta|_{\mathcal{E}_X}$, and the reduction of the internal Lagrangian is represented by any form $\varrho\in \Lambda^1(\mathcal{E}_X)$ such that
\begin{align*}
-d\varrho = -\,\tilde{\!\theta}^1_{xx} \wedge \,\tilde{\!\theta}^2 + \,\tilde{\!\theta}^1_{x} \wedge \,\tilde{\!\theta}^2_x - \,\tilde{\!\theta}^1 \wedge \,\tilde{\!\theta}^2_{xx} + 6(u^2 + v^2)\ \tilde{\!\theta}^1\wedge \,\tilde{\!\theta}^2,\qquad \,\tilde{\!\theta}^i_{kx} = \,\overline{\!\theta}^i_{kx}|_{\mathcal{E}_X}\,.
\end{align*}
For instance, one can take
\begin{align*}
\varrho =&\ u\, dv_{xx} - v\, du_{xx} - u_x dv_x + 6u^2v\, du - 6uv^2 dv + (u_x v_{xx} - v_x u_{xx}) \, dx \\
&+ \Big(-\dfrac{u_{xx}^2 + v_{xx}^2}{4} + (u^2 + v^2)\left(uu_{xx} + vv_{xx} - (u^2 + v^2)^2\right) + (uv_x - vu_x)^2\Big)dt\,.
\end{align*}

Let $\sigma\colon \mathbb{R}^2\to \mathcal{E}_X$ be a smooth section of the bundle $\pi_{\mathcal{E}_X}\colon (t, x, u, v, u_x, v_x, u_{xx}, v_{xx})\mapsto (t, x)$
\begin{align*}
\sigma\colon\quad
&u = a_0(t, x)\,,\ \ v = b_0(t, x)\,, \ \ u_x = a_1(t, x)\,,\ \ v_x = b_1(t, x)\,,\ \ u_{xx} = a_2(t, x)\,,\ \ v_{xx} = b_2(t, x)\,.
\end{align*}
Choose a compact submanifold $N^1\subset \mathbb{R}^2$. In Definition~\ref{Def1}, it suffices to consider paths of the form
\begin{align*}
\gamma(\tau)\colon\qquad
\begin{aligned}
&u = a_0 + \tau\delta a_0\,,\\
&v = b_0 + \tau\delta b_0\,,
\end{aligned}
\qquad
\begin{aligned}
&u_x = a_1 + \tau\delta a_1\,,\\
&v_x = b_1 + \tau\delta b_1\,,
\end{aligned}
\qquad
\begin{aligned}
&u_{xx} = a_2 + \tau\delta a_2\,,\\
&v_{xx} = b_2 + \tau\delta b_2\,,
\end{aligned}
\end{align*}
where $\delta a_i, \delta b_i\in C^{\infty}(\mathbb{R}^2)$ are arbitrary functions that vanish on $\partial N$. Then due to~\eqref{thehomotform},
\begin{align*}
&\dfrac{d}{d\tau}\Big|_{\tau = 0} \int_N \gamma(\tau)^*(\varrho) = \int_N \sigma^* (w \lrcorner\, d\varrho)\,,
\end{align*}
where $w = \delta a_0\, \partial_u + \delta b_0\, \partial_v + \delta a_1\, \partial_{u_x} + \delta b_1\, \partial_{v_x} + \delta a_2\, \partial_{u_{xx}} + \delta b_2\, \partial_{v_{xx}}$. One finds
\begin{align*}
&w \lrcorner\, d\varrho = - \big(\delta b_2 - 6(u^2 + v^2)\delta b_0\big)\,\tilde{\!\theta}^1 + \big(\delta a_2 - 6(u^2 + v^2)\delta a_0\big)\,\tilde{\!\theta}^2 + \delta b_1\ \tilde{\!\theta}^1_{x} - \delta a_1 \ \tilde{\!\theta}^2_x - \delta b_0 \ \tilde{\!\theta}^1_{xx} + \delta a_0 \ \tilde{\!\theta}^2_{xx}\,.
\end{align*}
At any point of $\mathcal{E}_X$, $w \lrcorner\, d\varrho = 0$ if and only if $w = 0$. Then $d\varrho$ defines the field of non-degenerate operators from $\pi_{\mathcal{E}_X}$-vertical vectors to Cartan $1$-forms. Hence, $\sigma$ is a stationary point of the reduction of the internal Lagrangian if and only if
\begin{align}
\sigma^* \,\tilde{\!\theta}^1 = \sigma^* \,\tilde{\!\theta}^2 = \sigma^* \,\tilde{\!\theta}^1_x = \sigma^* \,\tilde{\!\theta}^2_x = \sigma^* \,\tilde{\!\theta}^1_{xx} = \sigma^* \,\tilde{\!\theta}^2_{xx} = 0\,,
\label{Statpointsys}
\end{align}
i.e., if and only if $\sigma$ is a solution to $\pi_{\mathcal{E}_X}$ (that is $\sigma^*(\mathcal{C}\Lambda^1(\mathcal{E}_X)) = 0$). In terms of the components, system~\eqref{Statpointsys} consists of the equations
\begin{align*}
&\partial_x a_0 = a_1\,,\quad \partial_t a_0 = - \dfrac{b_2}{2} + (a_0^2 + b_0^2)b_0\,,\quad
\partial_x b_0 = b_1\,,\quad \partial_t b_0 = \dfrac{a_2}{2} - (a_0^2 + b_0^2)a_0\,,\\
&\partial_x a_1 = a_2\,,\quad \partial_x b_1 = b_2\,,\quad \partial_x a_2 = 6(a_0^2 + b_0^2)a_1\,,\quad \partial_x b_2 = 6(a_0^2 + b_0^2)b_1
\end{align*}
and their differential consequences
\begin{align*}
&\partial_t a_j = \partial_x^{\,j}\Big( - \dfrac{b_2}{2} + (a_0^2 + b_0^2)b_0\Big),\qquad \partial_t b_j = \partial_x^{\,j} \Big(\dfrac{a_2}{2} - (a_0^2 + b_0^2)a_0\Big),\qquad j = 1, 2\,.
\end{align*}
Thus the reduction of the stationary action principle for~\eqref{varprNLS} precisely reproduces $X$-invariant solutions of the NLS.

\vspace{1ex}

\remarka{\label{ThelastRem}
The NLS admits a $4$-dimensional commutative Lie algebra that is spanned by the Noether symmetries
$$
X,\qquad Y_1 = \partial_x,\qquad Y_2 = \partial_t,\qquad Y_3 = v\partial_u - u\partial_v + v_x\partial_{u_x} - u_x\partial_{v_x} + v_{xx}\partial_{u_{xx}} - u_{xx}\partial_{v_{xx}} + \ldots
$$
Then $\mathcal{E}_X$ inherits the symmetries $Y_1|_{\mathcal{E}_X}$, $Y_2|_{\mathcal{E}_X}$, $Y_3|_{\mathcal{E}_X}$. In accordance with the Noether theorem for invariant solutions (and Theorem~\ref{Theor3}), they give rise to the constants of $X$-invariant motion
\begin{align*}
&I_1 = u_x v_{xx} - v_x u_{xx}\,,\\
&I_2 = -\dfrac{u_{xx}^2 + v_{xx}^2}{4} + (u^2 + v^2)\left(uu_{xx} + vv_{xx} - (u^2 + v^2)^2\right) + (uv_x - vu_x)^2,\\
&I_3 = -uu_{xx} - vv_{xx} + \dfrac{3(u^2 + v^2)^2 + u_x^2 + v_x^2}{2},
\end{align*}
respectively. In simpler terms, for each $X$-invariant solution $\sigma$ of the NLS (global or local, with a connected domain $\subset \mathbb{R}^2$), there exist constants $C_1, C_2, C_3\in\mathbb{R}$ such that on the solution, $I_1 = C_1$, $I_2 = C_2$, $I_3 = C_3$ (more formally, $\sigma^*(I_i) = C_i$ for $i = 1, 2, 3$).

Note that $Y_i|_{\mathcal{E}_X}(I_j) = 0$ for $i,j = 1, 2, 3$. Hence $I_1$, $I_2$, $I_3$ are mutually Poisson commuting, where the Poisson bracket is determined by the inverse of $-d\varrho$ on fibers of~$\pi_{\mathcal{E}_X}$,
\begin{align*}
\{f, g\} = \mathcal{P}(df, dg)\,,\qquad
\mathcal{P} = \partial_{u_{xx}}\wedge\partial_v - \partial_{u_x}\wedge \partial_{v_x} + \partial_u\wedge\partial_{v_{xx}} + 6(u^2 + v^2)\partial_{u_{xx}}\wedge\partial_{v_{xx}}\,.
\end{align*}
Since $I_1$, $I_2$, $I_3$ are independent, one can informally say that $\mathcal{E}_X$ is Liouville integrable (even though it is not an ODE system), and its integrability is inherited from the NLS via invariant reduction.

The bundle $\pi_{\mathcal{E}_X}$ is trivial. Since
\begin{align*}
\{I_1,\ \} = \partial_x - D_x|_{\mathcal{E}_X}\qquad \text{and}\qquad \{I_2,\ \} = \partial_t - D_t|_{\mathcal{E}_X}\,,
\end{align*}
a function $f\in\mathcal{F}(\mathcal{E}_X)$ that does not depend on $t$ and $x$ is a constant of $X$-invariant motion if and only if
$$
\{I_1, f\} = \{I_2, f\} = 0\,.
$$
Thus $I_1$ and $I_2$ can be interpreted as Hamiltonians of two commuting vector fields that, together, reproduce~$\mathcal{E}_X$.
However, this interpretation is not invariant, and does not necessarily play a significant role in the integrability of $\mathcal{E}_X$.
}

}

\section{Conclusion}

The reduction framework proposed in this paper shows that equations satisfied by symmetry-invariant solutions inherit invariant geometric structures in a natural way. In this context, it does not matter whether the given local symmetry is a point or a higher symmetry. 

In some cases, finite-dimensional systems for invariant solutions of Lagrangian PDEs inherit integrability in the sense that the reduced presymplectic structures give rise to Poisson bivectors, and the reductions of conservation laws provide sufficiently many independent, mutually Poisson commuting constants of motion. This is the case in Example~\ref{ExampleNLS} (and also Example~\ref{ExamKB} after presymplectic reduction). At the same time, as Example~\ref{ExampleCBS} demonstrates, systems for invariant solutions may have additional geometric structures that cannot be obtained through the invariant reduction method. In both scenarios, finite-dimensional Liouville-integrable reductions of non-integrable systems with presymplectic structures may be of particular interest.

The presented reduction framework is global, which can be useful in studying qualitative (global) properties of symmetry-invariant solutions. In simple cases, it can be formulated as a computational algorithm. Examples~\ref{Examplenonldif},~\ref{Examplewave},~\ref{ExamKB}, and~\ref{ExampleNLS} demonstrate the algorithm in detail. A generalization of this algorithm to a broader class of equations is of practical importance.

Reductions of conservation laws, presymplectic structures, and internal Lagrangians of PDEs admit variational interpretations based on the homotopy formula. In various (multidimensional) situations, it may be possible to establish their non-triviality using properly defined stationary points. Other feasible approaches may involve, for example, multi-reduction.

The multi-reduction procedure described in Section \ref{Sectionmultired} is a natural consequence of our reduction approach. While some previously known approaches to multi-reduction such as the one by Anderson and Fels \cite{anderson1997symmetry} allow more efficient treatment of reduction of geometric structures such as Lagrangians and invariant representatives of conservation laws under suitable Lie groups of point symmetries, the multi-reduction in our framework does not have these limitations; in particular, it naturally allows the use of higher symmetries.

Since the reduction of presymplectic structures provides an alternative description of the reduction of variational principles, a general statement on the non-degeneracy of the reduction of non-degenerate variational principles is of considerable interest. Another natural step for future work is to describe the reduction of local Poisson brackets directly, without relying on presymplectic structures. To the best of the authors' knowledge, there is no known description of Poisson brackets in terms of the intrinsic geometry of PDEs in the general case, while the invariant reduction mechanism essentially relies on the intrinsic geometry. In its full generality, this problem promises to be challenging. The forthcoming third part of this study addresses this question for suitable classes of systems.

\section*{Acknowledgments }
The authors are grateful to NSERC of Canada for support through a Discovery grant RGPIN-2024-04308. K.D. is thankful to the Pacific Institute for the Mathematical Sciences for support through a PIMS Postdoctoral Fellowship.

\section*{Data availability}
Data sharing is not applicable to this article as no new data were created or analyzed in this study.

\section*{Conflict of interest} The authors have no conflicts to disclose.

\appendix

\section{Invariant reduction for PDEs without bundle structures}\label{App:A}

The reduction framework can be adapted for not necessarily regular PDEs without bundle structures. Note that, in the general case, if a vector field on $\mathcal{E}$ is the restriction of a vector field from $J^{\infty}(\pi)\supset \mathcal{E}$, its contraction with a differential $k$-form on $\mathcal{E}$ is well-defined.


If $Y$ is the restriction of a symmetry of $J^{\infty}(\pi)$ to an infinitely prolonged system $\mathcal{E}\subset J^{\infty}(\pi)$, and $\omega\in E_0^{\hspace{0.1ex}p, \hspace{0.2ex} q}(\mathcal{E})$ represents a $Y$-invariant element of $E_1^{\hspace{0.1ex}p, \hspace{0.2ex} q}(\mathcal{E})$, then there exists $\vartheta_{Y} \in E_0^{\hspace{0.1ex}p, \hspace{0.2ex} q-1}(\mathcal{E})$ such that $\mathcal{L}_Y\hspace{0.1ex} \omega = d_0 \vartheta_Y$. A reduction is represented by the well-defined restriction of $\vartheta_Y - Y \lrcorner\, \omega$ to the system $\mathcal{E}_Y\subset J^{\infty}(\pi)$ describing $Y$-invariant solutions, provided $Y$ restricts to it. Up to a sign, and in terms of the restriction $Y$ of a suitable point symmetry, the case where $\omega$ is $Y$-invariant and $\vartheta_Y = 0$ aligns with the reduction mechanism in Ref.~\cite{anderson1997symmetry}. If $Y$ is a trivial symmetry of $\mathcal{E}$ restricted from $J^{\infty}(\pi)$, then it coincides with the restriction of a trivial symmetry of $J^{\infty}(\pi)$, and one has $\mathcal{L}_Y \omega = d_0 (Y\lrcorner\, \omega)$.

\section{Generalization of Theorem~\ref{Theor1}}\label{App:B}

Theorem~\ref{Theor1} and the mechanism from Appendix~\ref{App:A} can be adapted to reduction of elements of the group $\widetilde{E}_1^{\hspace{0.1ex}0, \hspace{0.2ex} k}(\mathcal{E})$. Let us adopt the description~\eqref{canonisom}. Suppose $Y$ is the restriction of a symmetry of $J^{\infty}(\pi)$ to an infinitely prolonged system $\mathcal{E}\subset J^{\infty}(\pi)$. If $l\in \Lambda^{k+1}(\mathcal{E})$ represents a $Y$-invariant element $\boldsymbol \ell\in \widetilde{E}_1^{\hspace{0.1ex}0, \hspace{0.2ex} k}(\mathcal{E})$, then a reduction of $\boldsymbol \ell$ is determined by the restriction of $\vartheta_Y - Y\lrcorner\, l$ to $\mathcal{E}_Y$, where $\vartheta_Y\in \Lambda^k(\mathcal{E})$ is a differential form such that $\mathcal{L}_Y l - d\vartheta_Y\in \mathcal{C}^2\Lambda^{k+1}(\mathcal{E})$. We assume that $Y$ restricts to $\mathcal{E}_Y$. In this case, we have the following

\vspace{1ex}

\theorema{Suppose that $\widetilde{E}_1^{\hspace{0.1ex}0, \hspace{0.2ex} k-1}(\mathcal{E})|_{\mathcal{E}_Y} = E_1^{\hspace{0.1ex}2, \hspace{0.2ex} k-1}(\mathcal{E})|_{\mathcal{E}_Y} = 0$. If $\varrho\in\Lambda^{k}(\mathcal{E}_Y)$ represents the reduction of $\boldsymbol \ell \in \widetilde{E}_1^{\hspace{0.1ex}0, \hspace{0.2ex} k}(\mathcal{E})$, then $-d\varrho$ represents the reduction of $\tilde{d}_1 \boldsymbol \ell$.
}

\vspace{0.5ex}
\noindent
\textbf{Proof.} The reduction of $\boldsymbol \ell$ is well-defined due to $\widetilde{E}_1^{\hspace{0.15ex}0, \hspace{0.2ex} k-1}(\mathcal{E})|_{\mathcal{E}_Y} = 0$. Since $\boldsymbol \ell$ is $Y$-invariant, there are differential forms $\rho\in \Lambda^k(\mathcal{E})$ and $\mu\in \mathcal{C}^2\Lambda^{k+1}(\mathcal{E})$ such that
\begin{align*}
\mathcal{L}_Y l = d\rho + \mu\,.
\end{align*}
Then $\varrho = (\rho - Y\lrcorner\, l)|_{\mathcal{E}_Y}$ represents the reduction of $\boldsymbol \ell$,
and $\mu|_{\mathcal{E}_Y} = -d\varrho + (Y\lrcorner\, dl)|_{\mathcal{E}_Y}$. In addition,
$$
\mathcal{L}_Y (d l) = d\mu
$$
and hence the form
$\mu|_{\mathcal{E}_Y} - (Y\lrcorner\, dl)|_{\mathcal{E}_Y} = -d\varrho$ represents the reduction of $\tilde{d}_1 \boldsymbol \ell\in E_1^{\hspace{0.1ex}2, \hspace{0.2ex} k}(\mathcal{E})$.

\bibliographystyle{ieeetr}
{\small
\bibliography{References28}
}

\end{document}